
\input phyzzx

\def\Slacc{\centerline{\it Stanford Linear Accelerator Center}
\centerline{\it Stanford University, Stanford, CA 94309} }
\def\KKKK{\centerline{\it Department of Theoretical Physics,
University of Helsinki}
\centerline{\it Siltavuorenpenger 20 C, 00170 Helsinki, Finland } }
\def\SCIPP{\centerline {\it Santa Cruz Institute for Particle Physics}
\centerline {\it University of California, Santa Cruz, CA 95064}}
\def\LLLL{\centerline{\it Theoretical Physics Institute}
\centerline{\it School of Physics and Astronomy, 116 Church St. S. E. }
\centerline{\it University of Minnesota, Minneapolis, MN  55455 }}
\rightline{SLAC-PUB-5943}
\rightline{SCIPP 92/56}
\vskip 0.50in
\titlepage
\title{Hydrodynamic Stability Analysis of Burning Bubbles in Electroweak
Theory and in QCD \foot{Research supported by the Department of
Energy under grants DE-AC03-76SF00515 and DE-FG03-92ER40689 and by the
National Science Foundation under grant PHY89-04035.}}
\vskip 0.1in
\author{ Patrick Huet }
\address{\Slacc}
\author{ Keijo Kajantie }
\address{\KKKK}
\author{  Robert G. Leigh }
\address{\SCIPP}
\author{  Bao-Hua Liu and L. McLerran}
\address{\LLLL}
 \endpage \eject

\abstract{ Assuming that the electroweak and QCD phase transitions are
first order, upon supercooling, bubbles of the new phase appear. These
bubbles grow to macroscopic sizes compared to the natural scales
associated with the Compton wavelengths of particle excitations. They
propagate by burning the old phase into the new phase at the surface of
the bubble.  We study the hydrodynamic stability of the burning and
find that for the velocities of interest for cosmology in the
electroweak phase transition, the shape of the bubble wall is stable
under hydrodynamic perturbations. Bubbles formed in the cosmological
QCD phase transition are found to be a borderline case between
stability and instability.}

\endpage \eject

\def\ls#1{\ifmath{_{\lower1.5pt\hbox{$\scriptstyle #1$}}}}
\def\ie{{\it i.e.,}}%
\def\lsim{\raise0.3ex\hbox{$<$\kern-0.75em\raise-1.1ex\hbox{$\sim$}}}
\def\gsim{\raise0.3ex\hbox{$>$\kern-0.75em\raise-1.1ex\hbox{$\sim$}}}
\def\fr#1#2{{\textstyle{#1\over#2}}}
\def\\{\unskip\nobreak\hfill\break}
\font\subfont = cmss10
\font\ssfont =cmss10 scaled \magstep1
\def\q{\hbox{\ssfont q}}
\def\h{\hbox{\ssfont h}}
\def\subq#1{#1_{\lower1pt\hbox{\subfont q}}}
\def\subh#1{#1_{\lower1pt\hbox{\subfont h}}}
\def\vh{\subh{v}}       \def\vq{\subq{v}}
\def\gh{\subh{\gamma}}  \def\gq{\subq{\gamma}}
\def\wh{\subh{w}}       \def\wq{\subq{w}}
\def\ph{\subh{p}}       \def\pq{\subq{p}}
\def\Th{\subh{T}}       \def\Tq{\subq{T}}
\def\sh{\subh{s}}       \def\sq{\subq{s}}
\def\epsilonh{\subh{\epsilon}}
\def\epsilonq{\subq{\epsilon}}
\def\qq{\subq{q}}       \def\qh{\subh{q}}
\def\tmnq{\subq{\Theta}^{\mu\nu}}
\def\tmnh{\subh{\Theta}^{\mu\nu}}
\def\thetaq{\subq{\theta}}
\def\thetah{\subh{\theta}}
\def\Deltaq{\subq{\Delta}}
\def\Deltah{\subh{\Delta}}
\def\Sq{\subq{S}}       \def\Sh{\subh{S}}
\def\uq{\subq{u}^{\bot}}
\def\uh{\subh{u}^{\bot}}
\def\tem{\Theta}
\def\dtem{\delta\tem}
\def\k{{\rm k}}
\def\z{z}

\def\u{u^{\bot}}
\def\vb{v_{\rm b}} 
\def\wp{1+c_s^{-2}}
\def\Omeh{\widehat \Omega}
\def\db{d_{\rm b}} 
\def\dc{d_c} 



\REF\linde{D. A. Kirzhnits and A. Linde, {\sl Ann. Phys.}
 {\bf 101}, 195 (1976).}
\REF\anderson{D. Anderson and L. Hall, {\sl Phys. Rev.} {\bf D45}, 2682
(1992).}
\REF\carrington{M. Carrington, {\sl Phys. Rev.} {\bf D45}, 2933 (1992).}
\REF\lindea{M. Dine, R.G. Leigh, P. Huet, A. Linde, and D. Linde,
{\sl Phys. Lett.} {\bf B283}, 319 (1992).}

\REF\shaposhnikov{M.E. Shaposhnikov, JETP Lett. {\bf 44}, 465 (1986);\\
{\sl Nucl. Phys.} {\bf B287}, 757 (1987);  {\bf B299}, 797 (1988);\\
A.I. Bochkarev, S.Yu. Khlebnikov and M.E. Shaposhnikov,\\
{\sl Nucl.~Phys.} {\bf B329}, 493 (1990).}
\REF\mclerran{L. McLerran, {\sl Phys. Rev. Lett.} {\bf 62}, 1075 (1989).}
\REF\turok{N. Turok and J. Zadrozny, {\sl Phys. Rev. Lett.}
{\bf 65}, 2331 (1990); {\sl Nuc. Phys.} {\bf B358}, 471 (1991).}
\REF\mstv{L. McLerran, M. Shaposhnikov, N. Turok and M. Voloshin,
{\sl Phys. Lett.} {\bf B256}, 561 (1991).}
\REF\dine{M. Dine, P. Huet, R. Singleton and L. Susskind, {\sl Phys. Lett.}
{\bf B257}, 351 (1991).}
\REF\sakh{A.D. Sakharov, {\sl Pisma ZhETF} {\bf 5}, 32 (1967).}
\REF\thooft{G. 't Hooft, {\sl Phy. Rev. Lett.} {\bf 37}, 8 (1976);
{\sl Phys. Rev.} {\bf D14}, 3432, (1976).}
\REF\lindeb{A. Linde, {\sl Phys. Lett.} {\bf 70B}, 306 (1977).}
\REF\kuzmin{V.A. Kuzmin, V.A. Rubakov and M.E. Shaposhnikov, {\sl Phys. Lett.}
{\bf B155}, 36 (1985).}
\REF\arnold{P. Arnold and L. McLerran, {\sl Phys. Rev.}
{\bf D36}, 581, (1987);  {\bf D37}, 1020 (1988).}
\REF\carson{L. Carson, Xu Li, L. Mclerran and R. T. Wang,
{\sl Phys. Rev.} {\bf D42}, 2127 (1990).}
\REF\ambjorna{J. Ambjorn, M. Laursen and M. Shaposhnikov,
{\sl Phys. Lett.} {\bf 197B}, 49, (1989);
{\sl Nucl. Phys.} {\bf B316}, 483 (1990);
J. Ambjorn, T. Askgaard, H. Porter and M. Shaposhnikov,
{\sl Phys. Lett.} {\bf B244}, 479 (1990);
{\sl Nucl. Phys.} {\bf B353}, 346 (1991).}
\REF\qcd{For a review see for example D. Toussaint,
{\sl Nucl. Phys.} B[Proc. Suppl.] {\bf 26}, 3 (1992).}
\REF\wilczek{F. Wilczek, {\sl Int. J. Mod. Phys.}
{\bf A7}, 3911 (1992); R. Pisarski
and F. Wilczek, {\sl Phys. Rev.} {\bf D29}, 338 (1984).}
\REF\stein{ P. Steinhardt, {\sl Phys. Rev.} {\bf D25}, 2074 (1982).}
\REF\gyul {M. Gyulassy, K. Kajantie, H. Kurki-Suonio and
 L. McLerran, {\sl Nucl. Phys.} {\bf B237}, 477 (1984).}
\REF\hannu{H. Kurki-Suonio, {\sl Nucl. Phys.} {\bf B255}, 231 (1985).}
\REF\enqvist{K. Enqvist, J. Ignatius, K. Kajantie and K. Rummukainen,
{\sl Phys. Rev.} {\bf D45}, 3415 (1992).}
\REF\dlhll {M. Dine, R. G. Leigh, P. Huet, A. Linde and D. Linde,
 {\sl Phys. Rev.} {\bf D46}, 550 (1992).}
\REF\lmt {B. H. Liu, L. McLerran and N. Turok, TPI-MINN-92-18-T.}
\REF\keijo {K. Kajantie, {\sl Phys. Lett.} {\bf 284B}, 331 (1992).}
\REF\link {B. Link,  {\sl Phys. Rev. Lett.} {\bf 68}, 2425 (1992).}
\REF\landau {L. Landau, {\sl Acta Physicochimica} URSS,
{\bf XIX}, No. 1 (1944);
 L. D. Landau and E. M. Lifshitz, {\sl Fluid Mechanics}, \S 128,
(Pergamon Press,  Oxford, 1987).}
\REF\konto {V. M. Kontorovich, {\sl Soviet Physics} JETP {\bf 34},
 127 (1958).}
\REF\kkr {K. Kajantie, L. K\"arkk\"ainen and K. Rummukainen,
{\sl Nucl. Phys.} {\bf B357}, 693 (1991).}
\REF\potvin {S. Huang, J. Potvin, C. Rebbi and S. Sanielevici, {\sl Phys.
Rev.} {\bf D42}, 2864 (1990).}
\REF\freese{P. Kamionkowski and K. Freese, U.C. Berkeley Preprint
CFPA-TH-92-17.}
\REF\freeseneutrino{K. Freese and F. Adams, {\sl Phys. Rev.} {\bf D 41},
2449 (1990).}

\section{\bf Introduction}

There is renewed interest in the dynamics of the electroweak
phase transition\refmark{\linde~-~\lindea}.
This is due in part to the possibility that the
baryon asymmetry of the universe might be generated at such a
transition\refmark{\shaposhnikov-\dine}.
If the phase transition is first order
then all of the conditions necessary for generating a baryon excess
are present in the electroweak theory and its generalizations.

The Sakharov conditions for generating a baryon number excess
are\refmark\sakh:
\pointbegin Baryon Number Violation
\point  CP Violation
\point  Lack of Thermodynamic Equilibrium

The first condition is now well established in electroweak
theory\refmark{\thooft-\ambjorna}.
The rate of baryon number violation at temperatures larger than $T \ge$
100 GeV is known to be much greater than the expansion rate of the
universe, and not much less than a typical particle scattering time.

The second condition is satisfied in the standard model, although explicit
calculations show that the effects associated with
CP violation arising from the quark mass matrix
is much too small to generate an acceptably large
baryon asymmetry.  This is straightforward to
patch up in generalizations of the standard model by allowing CP
violation in the scalar interactions in an extended Higgs sector.

The third condition is satisfied if the electroweak phase
transition is first order.
The transition to the ordered low temperature phase
takes place by bubble nucleation.  Since all of the old phase
must pass through the walls of a bubble where the matter is strongly
out of thermal equilibrium, {\em all} of the matter is strongly
out of equilibrium
at some time during the phase transition.  If the baryon asymmetry is
made near or at the bubble walls, then all of the Sakharov conditions
can be satisfied.

The electroweak phase transition is believed to be of first order.
This has been shown to be true to the first few orders in perturbation
theory\refmark{\anderson-\lindea}. In higher orders, if $\alpha_W$
is small enough, then
these calculations are reliable.  For realistic values of
$\alpha_W \sim 1/30$, these results are somewhat controversial,
and perhaps may only be definitely resolved using non-perturbative
lattice methods. In this paper we will assume that the electroweak
phase transition is of first order, and will accept at face value
the results of perturbative computations.

If the QCD hadronization phase transition is of first order, there might
be consequences in both cosmology and heavy ion collisions.  Whether
or not the transition is first order is the subject of much
study using lattice non-perturbative methods\refmark\qcd.
The order of the transition depends sensitively on the number of quark
flavors and the quark masses and the physical case seems to be at the
borderline between first and second order transitions\refmark\wilczek.
In the following, we shall assume that the QCD phase transition is
first order.

To proceed with a semi-quantitative analysis of the effects
of a first order phase transition either in cosmology or during a heavy
ion collision, one needs to understand the formation and growth
of phase transition bubbles.  For the electroweak and QCD cases,
much is now known\refmark{\stein-\keijo}.
In particular,  the limiting velocity
of such walls has been computed for electroweak theory,
and the shape of the wall near the
burning surface is known.  In the QCD case, the macroscopic
features of the bubbles have been studied.

In this paper, we discuss the hydrodynamic stability of the
bubble propagation.  This has been discussed previously for
the QCD phase transition\refmark\link.
We find however that the previous analysis
was incomplete.   If this analysis is applied to either
QCD or electroweak theory, one would conclude that whenever the
wall velocity is less than the sound velocity of the matter,
then the bubble wall propagation is unstable.  Our results are
in contradiction with these conclusions.  The source of the discrepancy
is a more thorough treatment of the boundary conditions for the
motion of the bubble wall.

The outline of this paper is the following. In the second section of
this paper, we derive the macroscopic features of phase transition
bubbles for the electroweak phase in the 1+1 dimensional approximation,
extending known results for the case of QCD\refmark{\gyul-\hannu}. In
the third section, we derive the equations which describe a possible
convective instability which could affect bubble wall propagation.  Our
analysis is fully relativistic, but we also discuss the
non-relativistic limit.  We point out the origin of the discrepancy
between our results and those which have been previously derived. In
the fourth section, we solve for the possible instability for the
electroweak and QCD transitions.  For the electroweak theory, we find
stable growth for all velocities larger than
$\sqrt{\alpha_W/2\pi}\,\sim\, 0.07$. Detailed
computations\refmark{\dlhll,\lmt} of velocities favor $v > 0.1$, and so
this instability is of no physical interest. For the quark-hadron phase
transition, our analysis is inconclusive. A conclusive statement can be
made only after the relevant parameters of the phase transition are
known more accurately.  In the last section, we summarize our results.

\section{\bf Macroscopic Bubble Features}

Suppose we have a system which has a first order phase transition
at a temperature $T_c$.  If the system is expanding, then the system
supercools below $T_c$ before nucleating droplets of the low temperature
phase.  After these droplets have formed, they begin to grow
and expand into the system.  If the supercooling is sufficiently strong,
as is the case in the electroweak phase transition, then the droplets
expand until they collide with one another and complete the phase
transition. If the supercooling is sufficiently weak, as is the
case in the QCD phase transition in cosmology, then
the droplets expand until there is a mixed phase of droplets
of the low temperature phase embedded within the high temperature
phase.  The subsequent evolution of the mixed phase involves
coalescence of droplets, and as the system expands, an
increase in the volume of the low temperature phase at a rate determined
only by the rate of expansion of the system.  Eventually, all
of the system falls into the low temperature phase when
expansion has reduced the energy density of the system to that of the
low temperature phase.

In either case, for some time after nucleation, the bubbles
freely expand into the system eating up the high temperature phase.
If the growth of the bubbles is hydrodynamically stable, then
we expect that the bubble shape is smooth over macroscopic
distances and should be spherical.  The microscopic
physics is only important in a region of order the diffusion
length, a typical particle mean free path, near the surface of the
bubble wall.  This is seen explicitly in the hydrodynamic
equations, which admit similarity solutions, that is, where the
typical size scale is proportional to the time\refmark\gyul.

The analysis of a realistic situation is however a little complicated.
There are two cases one must distinguish, referred to as {\em
deflagrations} and {\em detonations}.

A detonation occurs when the velocity of the burning
front is larger than the sound velocity of the matter in the
high temperature phase.  This turns out to be equivalent to the
case when in the rest frame of the burning front, the velocity
of matter behind the burning front is less than that in front of it.

This burning front in its rest frame is shown in Fig. 1.
This is a frame which
we will use for analyzing possible convective instabilities.  For
understanding the physics, it is sometimes more useful to
consider the reference frame where the bubble wall moves, but
the matter far in front of it is at rest.

For the single front shown in Fig. 2  we have four unknowns, all other
thermodynamic quantities follow from the equation of state, $p=p(T),
s=p'(T),\epsilon(T)=Ts(T)-p(T)$. We shall neglect all chemical potentials.
 Between the four unknowns, two equations follow from conservation of
energy and momentum fluxes. By requiring that
$\partial_\mu \Theta^{\mu\nu}=0$ with\foot{Here $u^\mu$
is the fluid four-velocity
satisfying $u^2 = 1$. Our metric has signature
$(1,-1,-1,-1)$.} $\Theta^{\mu \nu} = (\epsilon + p ) u^\mu u^\nu - p g^{\mu
\nu}
   $,
 it follows that
$
\Delta\Theta^{\mu z}=0,
$
where $z$ is the direction of motion of the bubble wall. From these one
can, for instance, solve for the two velocities in terms of the two
temperatures
    as
$$
\vh^2 ={(\ph-\pq)(\epsilonq+\ph) \over (\epsilonh-\epsilonq)
(\epsilonh+\pq)}, \qquad \vq^2=
{(\pq-\ph)(\epsilonh+\pq) \over (\epsilonq-\epsilonh)(\epsilonq+\ph)}.
\eqn\velocities
$$
The system is then completely determined by firstly giving the temperature
$\Tq$
 of the supercooled \q-phase into which the bubble is
expanding. This is obtained from a nucleation calculation. Secondly,
a microscopic calculation\refmark{\dlhll-\keijo} gives the velocity
$\vb(\Tq)$ of the combustion front. This has so far been calculated for the
case
of $\vh-\vq\ll\vq$.

Boundary conditions often demand initial or final matter (or both)
to be at rest and the single front discussed above is not
sufficient. This leads one to a study of bubbles\refmark{\gyul-\hannu}.
To understand a detonation bubble it is useful to introduce space-time
rapidity,
$$
y = {1 \over 2} \ln\left( {{t+x} \over {t-x}} \right)
\eqn\defy
$$
and flow rapidity
$$
\Theta = {1 \over 2} \ln\left( {{1+v} \over {1-v}} \right).
\eqn\defTheta
$$
A 1-time + 1-space-dimensional detonation bubble is shown in Fig. 2, together
 with a plot of the flow rapidity and
energy density as a function of space-time rapidity.
Notice that constant space-time rapidity corresponds to $x/t \sim$ constant,
that is, to similarity growth.  It is well known that the
hydrodynamic equations admit similarity solutions.
Since the front is traveling faster than sound velocity,
the matter in ahead of the front is
unaffected by the front until the front actually comes in contact
with the matter.
Behind the burning front is fast
moving compressed matter, which over some region of rapidity has
a constant flow velocity and energy density.  The flow rapidity
must be determined by solving the hydrodynamic equations.  Behind this
region of constant flow velocity is a similarity rarefaction wave at
which the flow rapidity goes as
$\Theta(y) = y - y_s$, where $y_s=\tanh c_s$ is the rapidity associated
with the sound velocity $c_s$.  The flow rapidity can vanish when
the space-time rapidity is the rapidity of sound.  Therefore this
point moves with the sound velocity and tries to catch up with the
front.  It can never do so, because the front is moving supersonically.
The point at which the similarity rarefaction wave begins is when
$y = y_1 =y_{\rm flow}+y_s$.
In the very center of the bubble, the fluid
velocity must be zero.

A deflagration bubble is shown in Fig. 3.  A deflagration  (slow burning)
bubble propagates into the high temperature phase at a velocity less
than the sound velocity of the high temperature phase.  This turns out
to be equivalent, in the rest frame of the burning front,
to having the velocity of the fluid behind the front larger than
that ahead of it.  In this case,
the matter in front of the bubble is compressed and accelerated to a finite
velocity.  In front of the burning front, the moving matter produces
a shock which moves faster than the sound velocity.  (This
is analogous to the shock front produced by moving a piston down
a pipe).  Inside the burning front, the matter is at rest and at a
constant density.  Outside the shock front the matter is at rest and at
constant density.  The distance between the shock front and the
burning front increases as a function of time.

To compute the properties of these detonation and deflagration bubbles,
one must solve the hydrodynamic equations.  In the regions between
but excluding the fronts, there are acceptable solutions of the hydrodynamic
equations.  For the deflagration bubbles, the region excluding the front
is just a fluid moving at constant velocity, which will
solve the hydrodynamic equations.  For the detonation bubbles, the
rarefaction region has a solution which we will later explicitly
compute for the electroweak phase transition.

Before explicitly solving the bubble equations, it is useful to review
the properties of the solutions.  For deflagrations, we know the
energy density and velocity of the fluid to the right of the shock front.
We know the velocity of the fluid inside the burning front.  We must determine
the velocity of the burning front and the shock front, the energy density
inside the burning front, and the velocity and energy density
of the fluid in the region between the shock front and the burning front.
These are 5 unknowns.  The equations for the continuity of energy and
momentum flux across each front gives four equations.  To solve the problem
we need one more equation which is microscopic.  In electroweak theory,
this is the equation for the Higgs field.

For detonations, we know the velocity and energy density of the fluid
to the right of the burning front.  The energy density and velocity
of the fluid to the left of the burning front and the velocity of
the burning front are the 3 unknowns.  The continuity of energy
and momentum flux are 2 equations, and again we need a microscopic equation.

The bubble solutions have been explicitly discussed in Ref.~\gyul\
for the $1+1$-dimensional case and in Ref.~$\hannu$
for $1+d$, $d=1,2,3$.
We will now analyze these equations for the case of the electroweak
phase transition.

{\subsection {\it Relevant Scales and Parameters}}

 To proceed further, we must introduce the relevant physical quantities
and identify the scales and parameters which are the most useful to the
subsequent analysis. In the symmetric high temperature
 phase, we have
$$
\pq  = a\Tq^4, \qquad
\epsilonq = 3a\Tq^4.    \eqn\eosq
$$
In the symmetry broken phase, we have
$$
\ph = a\Th^4 \left[ 1 +  {\overline L}\;
\left(1-{\Th^2\over T_c^2}\right)
{T_c^4\over T_h^4}\right], \qquad
\epsilonh = 3a\Th^4 \left[ 1 - {{\overline L} \over 3}
\left(1 +{\Th^2\over T^2_c}\right) {T_c^4\over T_h^4}\right]\ .
\eqn\eosh
$$
Here,
$$
a =g^* {\pi^2 \over{90}}
$$
where $g^*\approx 107$ is the number of particle degrees of freedom. The
quantity
 $T_c$ is the critical temperature at which the
low temperature and high temperature phase have equal free energy,
that is, the phase coexistence temperature. We have introduced the
dimensionless parameter $\overline L$ which measures the
fraction of energy released in latent heat at the phase transition
$$
{\overline L}\,=\, {L \over  2 a T_c^4} \, =\,{3 \over 2} \
{\epsilonq(T_c) - \epsilonh(T_c) \over \epsilonq(T_c) }.
\eqn\Lbar
$$
${\overline L}$ is also a measure of the efficiency of the plasma to
absorb the latent heat released at the interface; on this ground, we expect
that
$$
\left| {\Tq-\Th \over \Tq}\right|\, \sim\, {\overline L}\,. \eqn\TqTh
$$
The important feature of the electroweak case is the intrinsic
weakness of the first order phase transition due to the smallness of
the parameter $\overline L$; typically ${\overline L}\leq 0.01$. In
particular, we can analytically solve for the structure of the
deflagration and detonation bubbles.

A second important parameter is the relative velocity of the front with
respect to the plasma, $\vb$, which can be written in the form
$$
{\gamma v}_{b}(\Tq) = {{V(\Tq)}\over {\cal E}} \eqn\vbb
$$
where $V(\Tq)\, = \, L\, ( 1 - {\Tq/T_c})$.
The function ${\cal E}$ is a slowly varying function of temperature,
which depends on the details such as mean free paths, numbers of
particles relevant at the temperature of the transition, \etc\ For our
purposes, we only need to know the form of $V(T)$, and in
addition, we need to know that computations of the numerical factors in
the above equation, for values of $\Tq$ relevant for cosmology, give for
the EW case ${\gamma v}_{b} \sim 1$ with velocities typically in the
range $0.1 \le\vb\le 0.9$. This range reflects the dependence on the
parameters of the standard model as well as the uncertainties which
result in its determination\refmark{\dlhll,\lmt}. For the QCD case
smaller values, $v_b \approx 0.04$, are indicated\refmark\keijo.
Unfortunately, more accuracy  is at present lacking
for determining the burning front velocity.

There are three scales to be considered in the stability analysis of
the interface to follow, and are denoted $\db$, $\dc$ and $\ell$.

$\bullet$ $\db$ is a time scale characteristic of the dynamics driving
the motion of the interface, defined as
$$
\db = {\sigma \over V(\Tq)} .\eqn\scaledb
$$
In the electroweak case, as shown in Section 4,
$d_b$ can be expressed in
terms of the thickness of the interface $\delta$ and a parameter
$\varepsilon$ ($0 < \varepsilon < 1$) proportional to the
amount of supercooling in the system, as
$\db\simeq\delta/\varepsilon$.

$\bullet$ $\dc$ is the scale
over which the surface tension $\sigma$ just balances the difference of
pressure across the interface, that is,
$$
\dc = {\sigma \over \pq-\ph}\ . \eqn\scaledc
$$
Its critical role in the stability of the interface was first stressed
by Landau\refmark\landau. In Section 4, it is shown that ${\dc /
\db} \, \ll \,1$ for the electroweak case, an inequality that we will
use later.

$\bullet$ $\ell$ is the mean free path of the particles which
contribute to the heat and momentum transfer at the interface. Its
effect for the electroweak transition is measured by the ratio $\ell
/ \delta$ which weights the damping of the wall motion in the plasma.
Because this ratio is typically less than 1, $\ell$ doesn't appear
in the following discussions. This is not the case at the QCD scale where
this ratio may be larger and can play a critical role in the stability
of the interface\refmark\freeseneutrino. These diffusive effects are
not considered here.

Having introduced the parameters ${\overline L}$, $\vb$ and the scales
$\db$, $\dc$ relevant to the problem and having set their relative
magnitudes, we may now proceed. We first turn to a detailed study of
the kinematics of deflagrations and detonations relevant to the
electroweak case. In the case of the QCD phase transition, the shapes
of the bubbles are much the same as is the case for the electroweak
phase transition.  The essential difference is that the latent heat of
the transition is larger, ${\overline L} \sim 0.5 - 1 $, and therefore
the discontinuities of energy densities are much larger.  Since
deflagration and detonation bubbles have been much discussed in the
literature for QCD, we shall not repeat this analysis here\refmark\gyul.

\subsection{ {\it Kinematics of Deflagrations}}

We first turn to the case of deflagrations. Using this parameterization
of the pressure and energy density, we can now solve for the velocity
of the shock front, $v_{\rm sh}$, the velocity of fluid in between the
shock front and the burning front, $v_{\rm flow}$, and the velocity of
the deflagration (burning) front $\vb$ denoted here by $v_{\rm def}$.
We define (see Fig.~3)
$$
x_0 =  {T_0^2 \over T_c^2},\;\;\;\;\; x_1 =  {T_1^2 \over T_c^2}
,\;\;\;\;\; x_2 =  {T_2^2 \over T_c^2}  \eqn\defofx
$$
with $T_2$ the
temperature to the right hand side of the shock front and the
temperature between the shock front and the burning front is $T_1$. In
the frame where the fluid is at rest inside the bubble, the results are
$$\eqalign{%
v_{\rm sh}   &= \sqrt{{3x_1^2 + x_2^2} \over {3(x_1^2+3x_2^2)}},
        \crr
v_{\rm flow} &= \sqrt{3} {x_1^2 - x_2^2 \over
\sqrt{(3x_1^2 + x_2^2)(x_1^2+3x_2^2)} }  \crr
             &= \sqrt{ {
\left[ x_0^2 - x_1^2 + \overline L (1-x_0)\right]
\left[3(x_0^2-x_1^2) - \overline L (1+x_0)\right]  \over
\left[3x_0^2 + x_1^2 - \overline L (1+x_0)\right]
\left[3x_1^2 + x_0^2 + \overline L (1-x_0)\right] }
}, \crr
v_{\rm def}  &= \sqrt{ {{ \left[x_0^2-x_1^2 + \overline L (1-x_0)\right]
\left[3x_1^2 + x_0^2 + \overline L (1-x_0)\right] } \over {
\left[3(x_0^2-x_1^2) - \overline L (1+x_0)\right]
\left[3x_0^2+x_1^2-\overline L (1+x_0)\right]
}}}.}\eqn\longequ
$$

When solving the above equations, we must require that the flux associated
with the entropy current
$$
        s^\mu = su^\mu
$$
where $s$ is the entropy density, be increased across the surfaces of
discontinuity.  These conditions are the same as
$$\eqalign{
      &x_1  \ge   x_2 \cr
     (x_0-x_1)^3 + \overline L \left(x_0 +x_1 + \right.
     &x_0x_1 -{\fr32} x_0^2  \left.
-{\fr32} x_1^2\right) - {\fr12}\overline L^2 (1-x_0)  \ge
0 \cr}\eqn\sameas
$$
When solving the above equations, we are given the parameters
$\overline L $ by knowing the latent heat at the phase transition,
$
        x_2 = {T_2^2 / T_c^2}
$
by knowing the amount of supercooling from the nucleation calculation,
and
$
        v_{\rm def}
$
by solving the microscopic equations for the burning front velocity.

For the weak deflagrations typical of the electroweak
phase transition, we can solve the equations explicitly.   We
have $T_2<T_c$ and assume
that $T_0-T_1 \approx T_2-T_1 \ll T_c$.
The regions where there are allowed solutions for the above
equations are then shown in Fig. 4. The linear boundaries of the
regions, given in the figure, correspond to $\vh=\vq=0,\,\, \vh=\vq=1,
\,\,\Delta s_\perp=0$\refmark\enqvist. The temperature
inside the bubble is
$$
  {T_0 - T_2 \over T_2} = {{\overline L} \over 6}\  {\sqrt3 v_{\rm def}
\over 1 + \sqrt3 v_{\rm def} }. \eqn\tzerottwo
$$
The temperature in the region between the shock front and the
deflagration front is
$$
        {T_1 - T_2\over T_2} = {{\overline L } \over 6}\
{{\sqrt{3} v_{\rm def}} \over {1 - 3 v^2_{\rm def}}}. \eqn\tonettwo
$$
Notice that $T_1 \ge T_0 \ge T_2$ and that (Fig. 4) $T_0<T_c$.  The flow
 velocity of the
matter between the deflagration front and the shock front is
$$
        v_{\rm flow} = {{3v_{\rm def}^2} \over {1 - 3 v_{\rm def}^2}}.
\eqn\vflowdef
$$
There is a singularity in these equations when
$v_{\rm def}^2 \rightarrow 1/3$, which
is an artifact of our approximate solution.  When
the burning velocity is very close to that of the sound velocity
in the supercooled phase, then the temperatures are not so
close to one another.

Since there is no reason for the burning velocity to be close
to the sound velocity, the solution above is sufficient for most
purposes.  For the weak first order electroweak phase transition,
the temperature is only slightly changed in the flow region and the
interior of the bubble wall.  The velocity of the fluid
is of order $v_{\rm def}^2$ which becomes of order 1 only for
deflagration velocities close to that of the sound velocity.

\subsection{ {\it Kinematics of Detonations}}

In the case of detonation bubbles, we have one surface of
discontinuity corresponding to the burning front.  Behind the
burning front, we must have a similarity rarefaction solution
of the hydrodynamic equations.  We assume that in the region of
the similarity rarefaction wave, the equation of state of the fluid is
close to that of an ideal gas.  In this case, we can solve
the hydrodynamic equations to find
$$
        \Theta = y - y_s
$$
and
$$
        \epsilon = \epsilon (y_s)\;\; {\rm e}^{4(y-y_s)/\sqrt{3}}. \eqn\epsy
$$

We can now solve for the velocities and energy densities in the
various regions subject to the constraint of positive entropy
generation.  In the region of small temperature
changes we find that
$$
        {\epsilon_1 - \epsilon_2 \over \epsilon_2} =
        {2\over 3}\ {{\overline L} \over {3v_{\rm det}^2 -1}}
\eqn\eoneetwo
$$
and
$$
        v_{\rm flow} = {\overline L \over 2}\
                        {v_{\rm det} \over {3v_{\rm
 det}^2 -1}}. \eqn\vflowdet
$$
The energy density far in the interior of the bubble is given by
$$
   {\epsilon_0 -\epsilon_2 \over \epsilon_2}=
   - {2 \over 3}\ {(1-\sqrt3)\overline L  \over 3v_{det}^2 -1}           .
 \eqn\ezeroetwo
$$
We therefore see that in the energy density in the interior of the
bubble is decreased relative to that in the outer region.  The matter
density in the flow region is compressed.  Again the changes are small
unless the burning front velocity is close to that of the sound
velocity.  In this case our analysis must be modified.

\section{\bf Stability of Relativistic Planar Combustion Fronts}

In this section we study the stability of planar
relativistic combustion fronts. The problem has been studied in the
non-relativistic case by Landau\refmark\landau\ and by Link\refmark\link\
for non-relativistic
velocities but for a relativistic equation of state.
The result is basically as follows.
Small perturbations of deflagration fronts are
stabilized by the interface tension $\sigma$ while large ones grow
exponentially with an initial growth time
 $$ \Omega^{-1}\simeq {4\over {\overline L^{\phantom L}} }\;{1\over \vb}\;
{\lambda^2 \over
\lambda -
 \lambda_c} \qquad \eqn\LL
 $$
where $\lambda_c$ is the critical length scale $\scaledc$ introduced in
the previous section
 $$
 \lambda_c = \dc\ . \qquad \eqn\lambdacLL
 $$
We would like to make two modifications to these earlier works.
First, in order to treat properly the critical role played by the
speed of sound $c_s\simeq {1 / \sqrt{3}}$,  the hydrodynamic
equations will be written in a completely
relativistic form. We can thus study the stability for fast
deflagrations, $\vh\rightarrow c_s$ and also the stability of detonations,
$\vq>\vh, \vq\ge c_s$. Secondly, we have now available several
microscopic estimates of the interface velocity $\vb$, at least for
the case $\vh-\vq\ll\vh$\refmark{\dlhll-\keijo}.
In particular, this velocity will
depend on the temperature $\Tq\approx\Th <T_c$. We shall see that if
this dependence is sufficiently strong -- as seems to be the case for
the electroweak theory -- the conclusions on the stability are completely
changed: the dependence of $\vb$ on $\Tq$ actually stabilizes large
scale perturbations of a deflagration burning front.

The stability analysis proceeds as follows. First the leading
hydrodynamic solution, a sharp planar combustion front separating a
region characterized by $\Tq,\vq$ from a region characterized by
$\Th,\vh$ is described\refmark\gyul. Secondly, the relativistic
hydrodynamic equations are linearized separately in the two regions
around a leading $T$ = constant, $v$ = constant solution (see also
Ref.~[\konto]) and the
linear equations so obtained are solved for the Fourier components of
the perturbations of $T$ and~$v$. The subtle aspect of the
analysis is then the joining of the solutions in the two regions
across a perturbed interface. The outcome of the stability analysis
depends on whether or not the connection equations have solutions with
$\Omega=-i\omega>0$; such solutions grow like $\exp(\Omega t)$ and
constitute an instability.

\subsection{\it Deflagrations and Detonations}

Here we review the results of Section 2 and formulate them in a form
more suitable to our purpose. The leading solution is defined as
follows. In the rest frame of the combustion front (assumed to be in
the $x,y$-plane), the velocities are parallel to the $z$-axis and the \q
(\h)-phase lies in the region $\z<0 \; (\z>0)$ (see Fig.~1). We assume
that the chemical potential $\mu=0$ and that the equations of state
$p~=~p(T)$ are
known. The configuration is thus entirely determined by the four
quantities $\Tq$, $\Th$, $\vq$, $\vh$. These are constrained by the
continuity of energy and momentum fluxes:
$$
\eqalign{
&\wq\gq^2\vq = \wh\gh^2\vh \equiv F_\epsilon \cr
&\wq\gq^2\vq^2+\pq=\wh\gh^2\vh^2+\ph \quad{\rm or}\quad
F_\epsilon(\vh-\vq)=\pq-\ph \cr }
\eqn\Tzeroth
$$where $w= p + \epsilon$ is the enthalpy of the fluid, and of entropy increase
$$
\sh\gh\vh \ge \sq\gq\vq. \eqn\fluxentropy
$$ Thus two of the four quantities remain unspecified by these general
relations.

Information on the further two required relations is obtained
from nucleation calculations and from microscopic calculations of the
wall velocity. The nucleation calculations give some temperature $\Tq
< T_c$ of the \q-phase, into which the bubbles of the \h-phase have to
expand. However, as described at length in Section 2,
boundary condition effects (matter at rest before and
after the transition) may require the \q-matter in front of the
interface to be shocked, \ie\ have a temperature higher than the
nucleation temperature. Similarly, a microscopic calculation of $\vb$
gives $\vh$ if boundary conditions demand the \h-matter behind the
front to be at rest or $\vq$ if the \q-matter in front of the front is
at rest. This distinction is not important if $\vh-\vq\ll\vh$. This is
the case for the EW transition, as we note from the constancy of
$F_\epsilon$ [Eq.~\Tzeroth] and the definition of $\overline L$
[Eq.~\Lbar], that
$$
{\vh -\vq \over\vq}\,\simeq\,{\wq(T_c) -\wh(T_c)\over \wh(T_c)}\, =\,
{{\overline L}\over
2}\,\leq \, 0.01. \eqn\vhvq
$$
 The reason why this velocity difference is so
small in the electroweak phase transition is because
the transition is rather weakly first order, most of the inertia
of the matter is carried by light mass degrees of freedom which are not
much affected by the transition and even the W,Z and Higgs bosons
gain a mass which is small compared to a typical kinetic energy
$E \sim 3T$.

\subsection{\it Linearized Hydrodynamic Equations}

We shall now consider either the \q- or the \h- region and linearize the
hydrodynamic equations $\partial_\mu \tem^{\mu\nu}=0$, $\tem^{\mu\nu}=
wu^\mu u^\nu -p g^{\mu\nu}$, $u^\mu=(\gamma,\gamma{\bf v})$, around a
solution $T$ = constant, ${\bf v}=(0,0,v)$ = constant. We shall use as
variables $\delta p=\delta w/(1+1/c_s^2)$, $\delta v$ and the
transverse 4-velocity variation $\u$. We assume that the sound velocity
is the same in the $\q$- and $\h$- phases. Instead of $v$ it is often
convenient to use the flow rapidity $\theta$ related to it by
$v=\tanh\theta, \delta\theta=\gamma^2\delta v$. Thus we write
$$
\eqalign{
p &=p_0+\delta p \cr u^\mu&=u_0^\mu +\delta
u^\mu=(\cosh\theta,0,0,\sinh\theta)+
(\sinh\theta\,\,\delta\theta,0,\u,\cosh\theta\,\,\delta\theta)
\cr } \eqn\lorentz $$ The components of the zeroth- and first-order
variations of the energy momentum tensor are then
$$
\eqalign{
\tem^{00} &= w\cosh^2\theta -p \cr
\tem^{0z} &= \fr12 w\sinh2\theta \cr
\tem^{0\bot} &=0 \cr \tem^{zz} &= w\sinh^2\theta +p \cr
\tem^{z\bot} &=0 \cr \tem^{\bot\bot} &=p \cr}\quad
\eqalign{
\dtem^{00}& =[(\wp)\cosh^2\theta-1]\delta p+w\sinh2\theta\,\,\delta\theta \cr
\dtem^{0z} &= (\wp)\fr12\sinh2\theta\,\,\delta p +w\cosh2\theta\,\,
\delta\theta \cr
\dtem^{0\bot} &= w\cosh\theta \,\, \u \cr
\dtem^{zz}& =[(\wp)\sinh^2\theta+1]\delta p +w\sinh2\theta\,\,\delta\theta
 \cr
\dtem^{z\bot} &= w\sinh\theta\,\, \u \cr
\dtem^{\bot\bot} &= \delta p  \cr} \eqn\stress
$$ The subscript 0 is omitted from $w_0$ and $v_0$ whenever no
confusion arises.

\def\papls{\partial_+}
\def\pamin{\partial_-}
\def\pabot{\partial_\bot}

The linearized equations $\partial_\mu\dtem^{\mu\nu}=0$ can then be
written in a particularly symmetric form by using light cone
derivatives $\partial_\pm = e^{\mp\theta} (\partial_0\pm\partial_z)$:
 $$
\eqalign{
&\fr12(c_s^{-2}+1)\papls\delta p +
\fr12(c_s^{-2}-1) \pamin\delta p +
w\papls\delta\theta + w\pabot\u=0, \cr
&\fr12(c_s^{-2}-1) \papls\delta
p +\fr12(c_s^{-2}+1)\pamin\delta p -
w\pamin\delta\theta+w\pabot \u=0,\cr
&\pabot\delta p+ \fr12w(\papls+\pamin)\u=0, \cr} \eqn\stresslin
 $$
Similarly, the first order variation of the entropy
conservation equation $ \partial_\mu \delta s^\mu= 0$ (valid
in the bulk fluid, far from the wall) becomes
 $$
\fr12 c_s^{-2}(\papls+\pamin)\delta p + \fr12 w
(\papls-\pamin)\delta\theta + w\pabot\u = 0.
\eqn\entropycont
$$

\subsection{\it Fourier Solution of the Linearized Hydrodynamic Equations}

We shall now search for solutions of Eqs.~$\stresslin$ of the form $$
\delta\theta=Ae^{-ik\cdot x},\qquad \u=B e^{-ik\cdot x},\qquad \delta p
= Ce^{-ik\cdot x}. \eqn\fourier $$ Since we are interested in
solutions behaving as $e^{\Omega t}$ but which do not blow up at large
distances,
it is convenient to introduce $\Omega =-i\omega$
and $q=ik_z$ and write the exponent in the form
 $$
 -ik\cdot x=
\Omega t + q\z+i \k x^ \bot. \eqn\eikx
 $$
The insertion of Eq.~$\fourier$ into Eq.~$\stresslin$ gives a linear
homogeneous system of equations for the Fourier components $A,B$ and $C$.
In general there is a dispersion relation
$$ {1\over
c_s^2}(\Omega+vq)^2 - (q+v\Omega)^2+{1\over\gamma^2}\k^2=0
\eqn\dispersion
$$
relating the components of the wavevector, and one can express
$\delta p$ and $\u$ in terms of $\delta\theta$. We find two types of
solutions:
 $$
\delta p= -w{\Omega+vq\over v\Omega+q}\delta\theta,\qquad
\u={i\k \over v\Omega+q}{1\over\gamma}\delta\theta. \eqn\generalsol $$
and
$$
\delta p=0,\qquad i\k\u ={-q\over\gamma}\delta\theta. \eqn\specialsol
$$ The latter is a special solution which describes an incompressible
velocity perturbation ($\nabla\cdot\delta{\bf v}=0$) moving with the
fluid, \ie\ it is only a function of $\z-vt,x^\bot$; it is
characterized by
$$
\Omega = -vq. \eqn\omegvq
$$
The dispersion relation, Eq.~$\dispersion$, deserves further attention;
on the $(\Omega,q)$ plane, it appears as the hyperbolae
$$ q_{\pm}={v(1-c_s^2)\Omega \pm
c_s(1-v^2)\sqrt{\Omega^2+(c_s^2-v^2)\gamma^2\k^2}
\over (c_s^2-v^2)} \eqn\qomegaplane
$$ with the asymptotes $$
\Omega={c_s-v \over 1-vc_s}q, \qquad \Omega=-{c_s+v\over 1+c_sv}q.
\eqn\asymptote
$$ It is plotted in Figs.~5. From the asymptotes one sees that for
$v<c_s$ both branches of the curve extend over all values of $\Omega$.
However, for $v>c_s$, one finds $q>0$ (actually
$q~\ge~\sqrt{1-v^2c_s^2}\gamma\k$) only for $\Omega<0$ and vice versa.
More generally, it is a simple exercise to show that the dispersion
relation $\dispersion$ has
no solution with $v \geq c_s \geq 0$, Re $q > 0$ and Re $\Omega > 0$. This fact
is of great importance for it establishes at once the stability of a detonation
front. Indeed, in Section 2, we have learned that a
detonation front is characterized by $\vq > c_s, \vh$.  On the other
hand, the perturbation behaves like exp$({\z\qq})$; as the \q-phase lies
in the region $\z \leq 0$,\foot{Physically imposed after the choice $\vq \geq
0.$} this requires Re $\qq \geq 0$. Hence, there is no possible unstable mode
(Re $\Omega > 0$ ) in a detonation front. In the remainder of our discussions,
we will assume that we are dealing with deflagration burning fronts only.

\subsection{\it Connecting the Solutions}

We now return to the two-phase situation depicted in the beginning of
this section and study how its small oscillations behave if the
interface is perturbed from $\z=0$ to
$$
\z=\zeta(t,x^\bot)= De^{\Omega
t+ i\k x^\bot}, \eqn\intfourier
 $$
where the amplitude $D$ of the perturbation satisfies
$$
\delta \ll D \ll 1/\k,\,1/\Omega. \eqn\require
$$
The condition that $  D \gg \delta $ is just the requirement that
we are looking for macroscopic fluctuations in the shape of the wall,
that is their characteristic size scale is much larger than the
thickness of the wall $\delta$.
The condition that $D \ll 1/k, 1/\Omega$ is the
requirement that we are looking at small fluctuations.
The solutions in the \q-phase ($\z<\zeta$) and in the \h-phase
($\z>\zeta$) can be immediately written down. Omitting a common factor
of $e^{\Omega t+i \k x^\bot}$ the solution in the \q-phase is
$$
\eqalign{\delta\thetaq &= A,\cr
\delta\pq &= -\wq {\Omega + \vq\qq \over \vq\Omega+\qq} A ,\cr
\subq{\u} &= {i\k \over \vq\Omega+\qq} {1\over\gq} A, \cr } \eqn\qphase
$$
where $\qq>0$ (to make the perturbation
vanish as $\z\rightarrow -\infty$). This requirement and the fact that we
are searching for $\Omega>0$ formally imply that one cannot include the
special solution $\specialsol$ with the dispersion relation $\Omega=-vq$ in
this region, as already emphasized by Landau\refmark\landau. However, in
the \h-region we must have $\qh<0$ and the special solution can (and must)
be included:
$$
\eqalign{\delta\thetah &= B+C,\cr
\delta\ph &= -\wh \;{\Omega + \vh\qh \over \vh\Omega+\qh}\; B ,\cr
\subh{\u} &=i\k \left[ {1 \over \vh\Omega+\qh} {1\over\gh} B -{\Omega
\over \k^2\vh\gh}C \right]. \cr } \eqn\hphase
$$
To see the physics of the special solution, one may compute
$$
({\bf \nabla}\times\delta{\bf \vh})^i = \epsilon_{i\bot z}\;
{\k\over\gh^2}
\left\{ {\vh\Omega \over \vh\Omega+\qh}B+
\left[1-\left({\Omega \over \k\vh}\right)^2\right] C \right\};
$$
in the \q-phase no $C$ term appears. As a consequence,
in the non-relativistic limit
$v\ll1$ this vanishes in the \q-phase and is nonzero in the \h-phase:
linear perturbations of the interface generate vortices which,
through the special solution, propagate in the \h-phase.

We thus have four
unknowns, $A$, $B$, $C$ and $D$ and need four equations to determine them
(or their ratios).
Three of the four conditions are obtained by expressing
conservation of energy-momentum across the interface $\partial_\mu
\tem^{\mu\nu}=0$ for $\nu~=~\bot,~0,~z$. Here, $\tem^{\mu\nu}$ is the total
stress-energy tensor of the system including the interface
 $$
  \tem^{\mu\nu}=\tmnq \theta\left[ -\z+\zeta(t,x^\bot)\right]
  + \tmnh \theta \left[ \z-\zeta(t,x^\bot)\right]
+\tem_K^{\mu\nu}\left\{\phi_K\left[\z-\zeta(t,x^\bot)\right]\right\} .
\eqn\Tfull
 $$
The two first terms are straightforward, with $\tmnq$ and $\tmnh$
constant.  The third term describes the stress-energy arising
from distorting the kink solution of the interface, assumed here to be
infinitely thin.
We have
$$\tem_K^{\mu\nu} = \partial^\mu\phi\partial^\nu\phi -g^{\mu\nu}
\left[\fr12 \partial_\lambda\phi \partial^\lambda\phi-V(\phi)\right]\ ,
$$
with the kink solution satisfying
$$
\phi_K''(\z)=V'(\phi_K), \qquad \fr12 \left[\phi_K'(\z)\right]^2
=V(\phi_K).
\eqn\kinfequation
$$
Calculating to first order in $\zeta$ one finds that
$$
\tem_K^{\mu\nu}= \left(
\matrix{
\sigma(\z)-\zeta\sigma'(\z) &0 &0 &\sigma(\z)\partial_0\zeta \cr
0 & -\sigma(\z)+\zeta\sigma'(\z) &0 &-\sigma(\z)\partial_x\zeta \cr
0 &0 &-\sigma(\z)+\zeta\sigma'(\z) &-\sigma(\z)\partial_y\zeta \cr
\sigma(\z)\partial_0\zeta & -\sigma(\z)\partial_x\zeta
 &-\sigma(\z)\partial_y\zeta &0 \cr }
\right),   \eqn\Tkink
$$
where the notation (for the surface tension)
$$
\sigma(\z)=\left[\phi_K'(\z)\right]^2
\approx \sigma\delta(\z-\zeta), \qquad
\sigma= \int d\z \sigma(\z), \eqn\sigmapprox
$$
has been introduced.

If we now compute $\partial_\mu\tem^{\mu\nu}$ from Eq.~$\Tfull$, the result
is proportional to $\delta(\z-\zeta)$. For the two first
terms this arises as the derivative of the $\theta$-function. For the
kink term, Eq.~$\Tkink$ gives $\partial_\mu \tem_K^{\mu\nu}=\sigma(\z)\;
\partial^2\zeta\; \delta^{\nu z}$, which again, within the approximation
$\sigmapprox$, is proportional to $\delta(\z-\zeta)$. A quantity proportional
to a $\delta$ function can only vanish if the coefficient of the
$\delta$ function vanishes. We thus obtain, at the wall:
$$\subq{\tem}^{z\nu}=\subh{\tem}^{z\nu}\eqn\eqstresso$$
$$
 \bigl[ \subq{\dtem}^{z\nu} -\subq{\tem}^{0\nu}
 \partial_0\zeta -\subq{\tem}^{\bot\nu}\pabot\zeta \bigr]
 - \bigl[ \q\rightarrow \h \bigr]
-\sigma(\partial^2_0-\partial^2_\bot ) \zeta\,\delta^{z\nu}=0\ .
\eqn\eqstress
$$
The first equations reproduce the leading conservation conditions
$\Tzeroth$.\foot{In general, there is a term on the RHS of Eq.~
\eqstresso\ involving the extrinsic curvature of the wall.} The others
give us three linear relations between the coefficients $A$, $B$, $C$,
and $D$.

A fourth equation can be obtained by expressing conservation of entropy
across the wall,
$$\gq\vq\sq=\gh\vh\sh + \int_{wall} J_{source}, \eqn\entropysource$$
where the last term on the right-hand side is a measure of the flux of
entropy gained by the plasma as it crosses the wall. From the work of
Refs.~[\dlhll--\keijo],
it has been established that entropy generation reflects
directly the existence of a velocity dependent force which damps the
wall motion to a terminal velocity. This velocity was discussed
in the second section,
$$
(\gamma v)_b = {V(\Tq) \over {\cal E}}\ . \eqn\vwall
 $$
 This equation of motion has to be modified to include the effect of
 the curvature of the interface as well as of its acceleration,
 namely,
 $$
 \sigma \,(\partial^2_0  - \pabot^2) \, \zeta =
 - V(\Tq) + (\gamma v)_b {\cal E}\ . \eqn\vwallmod
$$
Equation $\vwall$ yields a unique value for the entropy source in
$\entropysource$ and, consequently, can be alternatively used to
generate the needed fourth relation. After linearizing Eq.~$\vwallmod$,
we obtain
$${ \vq \db }\,(\partial^2_0 -
\pabot^2) \, \zeta = ( \delta\thetaq - \gq^2\partial_0 \zeta) -
  \eta\, \, \delta\thetaq \eqn\eqfour
$$
where $\db = {\sigma / V(\Tq)}$, a scale we
have already introduced in Eq.~$\scaledb$, and where one has defined
$$\eqalign{
\eta&= -\left(-T_c{d\vb\over d\Tq}\right)
{1\over \wq}\ {\delta \pq \over \delta \thetaq}\crr
    &=\left(-T_c{d\vb\over d\Tq}\right) {\Omega+\vq\qq
       \over \vq\Omega+\qq} \crr
    &\simeq
         \left(-T_c{d\vb\over d\Tq}\right) \vb . \cr}
\eqn\etaterm $$
The latter approximation will be justified later.

Equations $\eqstress$ and $\eqfour$ form a complete set of homogeneous linear
equations
$$ {1\over\gq} \uq + \vq\pabot\zeta
= {1\over\gh} \uh + \vh\pabot\zeta.
\eqn\eqone
$$
$$
\eqalign{
(\wp)\gq^2\vq\delta\pq
&+\wq\gq^2(1+\vq^2)(\delta\thetaq
   -\partial_0\zeta)  \cr
&=(\wp)\gh^2\vh\delta\ph
   +\wh\gh^2(1+\vh^2)(\delta\thetah -\partial_0\zeta).
\cr}\eqn\eqtwo
$$
$$
\eqalign{
\left[(\wp)\gq^2\vq^2+1\right] \delta\pq&+2\wq\gq^2\vq\delta\thetaq
-\sigma({\partial^2_0-\partial^2_\bot})\zeta  \cr
&=
\left[(\wp)\gh^2\vh^2+1\right] \delta\ph+2\wh\gh^2\vh\delta\thetah.
}\eqn\eqthree
$$
$$
\gq^2\partial_0 \zeta + {\vq \db }\,(\partial^2_0 -
\partial^2_\bot)\zeta = ( 1 - \eta) \, \,\delta\thetaq .\eqn\eqfourbis
$$
Eq.~$\eqone$ relates the changes of transverse velocity while Eqs.~$\eqtwo$ and
$\eqthree$ describe the conservation of the flow of energy and momentum
respectively. If $\eta$ were zero, Eq.~$\eqfourbis$ tells us
that the velocity $\vb$ of the interface relative to the \q-phase is
only perturbed by surface tension effects; this is akin to
the boundary condition initially used by Landau.\foot{In his analysis,
Landau further ignored the effects of the surface
tension. This is reasonable for most of the applications.}
However, due to the potential dependence of the interface velocity
on the temperature, this is no longer true and the $\eta$ term is
needed. Furthermore, Eq.~$\eqfourbis$ provides a direct physical
interpretation of the
stability of a deflagration that we will uncover in the next section.
We now proceed to the resolution of this system of equations.

\subsection{\it Solution for Small Velocities}

The case of general velocities results in rather lengthy equations
and we shall first discuss the solution for small velocities.
In this case, the equations are much simpler and their interpretation is
more transparent. This case is defined by $\vq,\vh\ll c_s$
for which, the integrability condition $\dispersion$ simply implies
$$
\qq \simeq \k,\qquad \qh \simeq -\k,\qquad |\omega| \sim v \k.
\eqn\NRqq
$$
As a result the solutions in the \q- and \h-phases simplify to
$$
\eqalign{
\delta\vq &=A \cr
\delta\pq &= -\wq \left({\Omega\over \k}+\vq\right)A \cr
\subq{\u} &= i A, \cr }\qquad
\eqalign{
\delta\vh &=B+C\cr
\delta\ph &= \wh\left({\Omega\over \k}-\vh\right)B \cr
\subh{\u} &= -i \left(B+{\Omega\over \vh \k}C\right).  \cr }
\eqn\NRfluct
$$
After inserting these expressions into Eqs.~\eqone--\eqfourbis\
and using $\Tzeroth$ and $\vhvq$ as well as the definitions of $\dc$, we obtain
$$
\eqalign{\pmatrix{1&1&{\Omega\over \k\vh}&-\vq{ {\overline L} \over 2}\cr
\vh& -\vq& -\vq &-\vq {{\overline L} \over 2}{\Omega\over \k}\cr
({\Omega\over \k\vq}-1) & ({\Omega\over \k\vh}+1) & 2 &
 \vq {{\overline L} \over 2} \dc \k\cr
-(1-\eta) & 0 & 0 & ({\Omega\over\k}+\vq \db \k)\cr}}\quad
\eqalign{\pmatrix{A\cr B\cr C\cr D\cr}}=0\ .
\eqn\NRsystem
$$
A non trivial solution exists if and only if the determinant of the
matrix in Eq.~$\NRsystem$ vanishes, that is,
$$
\Omega = \k\vh \eqn\spurious
$$ or
$$
\vq^2(1-\eta){{\overline L} \over 2} \left({\k\over k_c}-1\right)
+2 \vq \left[1+\eta{{\overline L} \over 2}
+ \db \k \right]  { \Omega \over \k}
+\left[1 + {\vq\over\vh}+{\vq\over\vh}\ \eta\
      {\overline L \over 2}\right]
\left({\Omega\over\k}\right)^2 = 0\ . \eqn\NRdet
$$
with
$$
\lambda_c = k_c^{-1} = \dc \, \left(1+ {2\over 1-\eta}
         \ {2 \over {\overline L^{\phantom l}}}\, {\db \over \dc}\,\right)
\eqn\lambdac
$$
defined as the critical wavelength.
The first solution $\spurious$ is spurious and has no physical
significance. It leads to $A=D=B+C=0$ and from Eq.~$\NRfluct$ the
entire solution vanishes. In general, it lies at
$\Omega=\gh\vh \k$, and the solution vanishes likewise. The smallness
of $\vq$ insures that the term quadratic in $\Omega$ is small and the
solution is
$$
{\Omega \over \k} \simeq{ {\overline L} \over 4}\,{\vb (1- \eta)
(1 - {k/k_c}) \over 1 + \eta\ \overline L/2
 + \db \k }. \eqn\NRsol
$$
For our purpose, Eq.~$\NRsol$ adequately generalizes the original Landau
 formula $\LL$.
 Accordingly, the condition of stability becomes
$$
(1-\eta)(\lambda -\lambda_c)<0. \eqn\stabcriterion
$$
If $\eta<1$ the system is unstable for perturbations of scale $\lambda$
larger than $\lambda_c$.
This is the case if the wall
velocity $\vq$ depends sufficiently weakly on $\Tq$. However, if
this dependence is strong enough to make $\eta>1$, {\it large scale
perturbations become stable}.

We can now comment on two numerical approximations made earlier.
Firstly, it is obvious from Eq.~$\NRsol$ that $\Omega$ is of the order
of ${\overline L}\vq \k$ so that indeed the relevant range of $\Omega$
is $\leq v \k$. Secondly, including the $\Omega$ dependence of $\eta$
in Eq.~\NRdet\ would amount to $\vq\rightarrow \vq+\Omega/\k$. However,
the additional (positive) term would not contribute to the
$\Omega$-independent terms and thus is irrelevant for the stability
consideration.

We now turn to the interpretation of the results
\lambdac--\stabcriterion. First, we note that the critical
wavelength $\lambda_c$ [Eq.~\lambdac] receives a correction of order
${\overline L}^{\; -1}$ in respect to Landau's value $\dc$
[Eq.~\lambdacLL]; this
term  originates from the RHS of Eq.~\vwallmod. In a typical
macroscopic situation, the latent heat released is a significant
fraction of the total energy available (${\overline L} \gg 1$) and this
correction is negligible. However, in the extremely weak phenomena
considered here (${\overline L} < 1$), this correction is large and actually
controls the value of $\lambda_c$.

$\bullet$ If $\eta \ll 1$, $\lambda_c \simeq (4/\overline L)\
\db$ and is significantly larger than $\dc$ because of the
smallness of ${\overline L}$ and of the ratio ${\dc/\db}$.

$\bullet$ If $\eta\, \gg\,1$,
$\lambda_c \simeq - (1/\eta)\ (4/\overline L)\ \db$
and is driven negative; in which case, according to Eq.~%
\stabcriterion, {\it perturbations at all scales are stable}.

The physics of this stability restoration depending on the size of
$\eta$ is as follows. For the sake of simplicity, let us ignore (with
Landau) the RHS of Eq.~\eqfour\ which doesn't play any role in
the argument. In the limit $\eta\to 0$, Eq.~\eqfour\ reproduces
Landau's boundary condition. In such a case, the velocity of the perturbation
$\partial_0 \zeta$ follows the variations in velocity of the fluid.
This drives perturbations larger and larger unless the surface energy
$\pabot^2 \zeta \sim {\zeta / \lambda^2}$ is significant enough to
prevent their growth; this occurs for $\lambda$ sufficiently small. In
the present case, the interface velocity [Eq.~\vwall]
is proportional to
$V(\Tq)$, an extremely sensitive function of the temperature of the
\q-phase; this sensitivity is measured by $\eta$. The growing perturbation
triggers a local variation in temperature\foot{Positive or negative
depending on its direction of growth.} which affects, in turn, the
velocity independent pressure $V(\Tq)$ acting on the interface. This
increase or decrease in pressure acts as a restoring force which opposes
the growth of the perturbation. When this force is large enough to
dominate the others ($\eta > 1$), the perturbation collapses. This
mechanism operates only if the characteristic growth time of the
perturbation is larger than the time scale which characterizes the wall
dynamics, that is, only if $\Omega \db < 1$. This condition is amply
satisfied in the cases of interests.\foot{As $\Omega \ll \k$ and $\k$ is
restricted by Eq. $\require$.}

At this point we may also compare with the 4th equation
used by Link\refmark\link\ which in the present notation is
$$
\Omega A= A + {3\vq \over 4\vh}\alpha
\left[\left({\Omega\over \k}+\vq\right)A
+\left({\Omega\over \k}-\vh\right)B\right], \eqn\link
$$
where $0\le\alpha\le1$ is a phenomenological parameter used to
describe the magnitude of $F_\epsilon$ in terms of $\Tq^4-\Th^4$.
It is evident that in this case the additional term is only a
small correction to $A$ and will not modify the conclusions about
stability. Our calculation, on the other hand, is based on an
explicit calculation of $F_\epsilon=\wq\gq^2\vq(\Tq)$ and indicates that
this dependence may be strong enough to actually make large scales
stable.

\subsection{\it Solution for Arbitrary Velocities}

As mentioned earlier, the speed of sound $c_s$ plays a critical
role in the stability of a moving front. In order to extend the
previous conclusions to velocities ranging up to $c_s$, one has to
generalize the analysis of the previous section.
It is convenient to introduce the dimensionless quantities
$$
\Omeh ={\Omega\over \gh\vh \k}
\eqn\hatomega
$$
$$
 \eqalign{
\Deltah &= \sqrt{ 1-{\vh^2\over c_s^2}}, \qquad \Sh(\Omeh) =
 \sqrt{\Omeh^2{\vh^2 \over c_s^2} + \Deltah^2} \cr
\Deltaq &= \sqrt{ 1-{\vq^2\over c_s^2}}, \qquad \Sq(\Omeh) =
 \sqrt{ \Omeh^2{\vh^2 \over c_s^2} + \Deltaq^2 { \gq^2 \over \gh^2}}
. \cr} \eqn\rescale
 $$
The calculation of the determinant proceeds exactly as before and we
will only quote the results here.
Neglecting once again the spurious solution $\Omeh=1$, we may write the
determinantal equation in the form:
$$
\Omeh\, {\cal D}(\Omeh) = {\cal N}( \Omeh ). \eqn\Rdet
$$
${\cal N}$ and ${\cal D}$ are two lengthy polynomials in
$\Omeh$, $\Sq$ and $\Sh$. ${\cal D}$ is
a positive quantity unless ${\k/k_c} $ is
excessively large,
while ${\cal N}$ contains the critical information on
the stability behavior we have uncovered in the small velocity limit.
A closer inspection of Eq.~$\Rdet$ reveals that the absolute value of
$\Omeh$ ($\, = |{{\cal N} / {\cal D} }|$) is bounded from
above and is never much larger than
one. Consequently, we can attempt to solve $\Rdet$ iteratively
in the following manner
$$
\Omeh_0= {{\cal N}(0) \over
{\cal D}(0)} \,, \qquad \Omeh_1= {{\cal N}(\Omeh_0) \over
{\cal D}(\Omeh_0)}\,, \qquad { \it etc}. \eqn\iteration
$$
This iteration converges very rapidly. A comparison of the lowest
order of this
approximation with a numerical analysis shows an agreement better than
one percent over the whole range of velocities. We then obtain
$$
{\Omega \over \k}\simeq\gh\, \vh\, {{\cal N}(0) \over
{\cal D}(0)} \eqn\Rsolution
$$ with
$${\cal N}(0)= (1-\eta)\, { {\overline L}\over 2}\,\gq\vq\,
\Deltaq \left(1-{\k\over k_c}\right)\,\Deltah $$
$$\eqalign{
{\cal D}(0)&\simeq \gh\vh \left(\Deltah+\Deltaq {\vh \over \vq}\right)
\gq^2 + \gq\vq \db \k \left(\Deltah+{\vh\over\vq}\Deltaq\right)
      \cr
  &\qquad + \eta {{\overline L} \over 2}
       \Biggl\{\Deltaq \gh\vh + \Deltah \gq\vq [\Deltaq
       + (2+\Deltaq^2)\gq\vq\gh\vh ] \Biggr. \cr
 &\hskip1.2cm  - \gq\vq \dc \k \left[ \Deltaq \gh^2 \vh^2  \left.
 \left(1 +{3\over 1 + \Deltah}\right) + \vq\vh  \gq^2 \gh^2 (2+\Deltaq^2)
 \left(1+\vh^2\right)\right]  \right\}\cr}
  \eqn\NDzero$$ and
$$
\lambda_c= k_c^{-1} = \dc \,
 { \gh ( 1 + \vh^2) \over \Deltah} + {2 \over 1-\eta}\ {1 \over
 {\overline L^{\phantom l}}}\,\db\,
 \left( {\vh \over \Deltah }+{\vq \over \Deltaq}\right)
 {1 \over \vq \gq}
  \eqn\Rlambdac
 $$
Solution $\Rsolution$, along with $\Rlambdac$, suitably generalizes
the corresponding expressions $\NRsol$ and $\lambdac$ for velocities ranging
up to $c_s$.
As a first check, the reader may verify that, as $\vq$ and $\vh$
approach zero, $\Sq$, $\Sh$, $\Deltah$ and $\Deltaq$ all approach
unity and Eqs. $\Rsolution$ and $\Rlambdac$
reproduce the familiar small velocity results. Furthermore, the condition for
stability of a disturbance reads
$$
 {\cal N }(\Omeh=0) \leq 0 \eqn\Rstability
$$
and it is clear that all the conclusions reached
previously hold. In particular, $\eta > 1$ stabilizes the long
wavelength modes.

As $\vh \rightarrow c_s$, the quantity $\Deltah\rightarrow 0$ and
expression $\Rlambdac$ indicates that $\lambda_c\rightarrow \infty$ in
this limit; in such  a case, our solution $\Rsolution$ is no longer
appropriate.\foot{Only for ${ \vh / c_s} > 0.99 $.} However, one
may show, using the full secular equation $\Rdet$, that, in this
limit, the determinant possesses no positive roots for any value of
$\k$. This result matches nicely with the stability of detonations.

\section{\bf Applications to the EW and QCD Phase Transitions}

In this section we apply the results of the previous sections to the
evolution of an expanding bubble of true vacuum produced at the
electroweak phase transition. The characteristics of the electroweak
phase transition are entirely encoded in the finite temperature
effective potential for the Higgs field
$$
        V(\phi,T) = D (T^2 - T_0^2) \phi^2 - E T \phi^3
+ {\lambda_T \over 4} \phi^4.  \eqn\effpot
$$

The parameters $D$ and $E$ may be
expressed in terms of the weak coupling constant
$\alpha_W = g^2/4\pi \sim 1/30$ as
$$
        D \approx {5 \over{32}} g^2, \qquad
        E \approx {g^3 \over{16\pi}}. \eqn\valueed
$$
Here we have used $m_W \approx m_Z \approx m_t$.   We can also express
the Higgs self-coupling as
$$
        \lambda_T = g^2 {m_H^2 \over 8m_W^2}, \eqn\valuelambda
$$
where $m_H$ is the Higgs mass. This is an effective potential valid
near the phase transition temperature where multiple scalar degrees of
freedom may be important, and therefore $m_H$ may not be the physical
Higgs particle mass\refmark\lindea. In order to generate a baryon
asymmetry which is not washed out in the broken phase, the
effective Higgs mass must be $m_H \lsim 40$ GeV.

In Eq.~\effpot, $T_0$ is the temperature at which the Higgs mass
vanishes and is the lowest temperature for which the high temperature
phase is metastable, $T_0~\sim~(m_H/m_W)\,~250$~GeV.
The temperature at which the phase transition nucleates
is within the range $T_0 < T < T_c$, where
$$\eqalign{
{T_c^2-T_0^2\over T_c^2}&= {E^2\over\lambda_T D} \cr
                       &\approx {\alpha_W \over \pi}{m^2_W \over
                        m^2_H}\cr
                       &\lsim \, 0.04\, . \cr} \eqn\tzerotc
$$
In terms of these parameters,
$$
\ph\, =\, a\Th^4\, -\, V(\Th) \, , \eqn\Vh
$$ where $V(\Th)$ is the value of $V(\phi,\Th) $ at its absolute minimum.
We can adequately express it as
$$
V(\Th) \,=\, \varepsilon \, V_c   \eqn\VTq
$$
with
$$
\eqalign{\varepsilon \, &= \,{T^2_c-\Tq^2 \over T^2_c -T^2_0}  \cr
                        &\propto \, \, \, {1\over m_H^{3/2}} \cr }
                           \qquad \, \, {\rm and} \, \qquad \,
\eqalign{V_c \, &= \,\,\, {1\over2}{E^4 \over \lambda^3}\, T_c^4 \cr
                &\simeq \, {2 \over  \pi}\, \alpha_W^3 {m_W^6 \over
m_H^6}\, T_c^4 \cr}\eqn\epsVc
$$
where $\varepsilon$ measures the amount of supercooling at the phase
transition; it has the
value $0.25$ for $m_H \, = \, 40$ GeV.

The wall thickness $\delta$ can conveniently be expressed in terms of
$V_c$ and the surface tension $\sigma =\, \int{d \phi \, \sqrt{2
V(\phi,T_c)} } $ as
$$
\eqalign{
\delta &= \int_{\it wall} {d \phi \over \sqrt{2 V(\phi,T_c)}} \cr
    &\simeq {\sigma \over V_c}
     \, \sim \,10 - 100\,\;\;T^{-1}\ . \cr } \eqn\thickness
$$
We can now develop an intuition for the scale $\db$ ($ = {\sigma
/ V(\Tq)}$) which appears naturally in our calculations; from its
definition and Eqs.~$\VTq$ and $\thickness$, we can express it as
$$
\db \, \simeq \, {\delta
\over \varepsilon} \,. \eqn\dbdelta
$$
In the situation of maximal supercooling ($\varepsilon\, = \, 1$)
$\, \db \, = \delta$ and it increases as the phase transition becomes more
weakly first order ($\varepsilon\, \rightarrow \, 0$).

According to Eqs. $\eosh$ and $\Vh$, the relative latent heat is
$$
\eqalign{
        \overline L  &=  {8 \over{2a}}
{E^2  D  \over  \lambda_T^2 } \left(1-  {E^2 \over
{\lambda_T D}}\right) \cr
              &\approx {5 \over {2a}} \alpha_W^2 { m_W^4 \over m_H^4}
              \cr
             &\lsim \, 0.01 \, . \cr} \eqn\Latentheat
$$
In the case of $\eta > 1$, we have asserted that there are no unstable
perturbations on the basis that perturbations of size larger than the
critical wavelength $\lambda_c$ are stable and that $\lambda_c$ has the
negative value
$$
\lambda_c \simeq -\, {1\over\eta}\
{4\over{\overline L^{\phantom l}}}\ \db\, .
\eqn\newlambdac
$$
{}From Eq.~$\lambdac$, this is a trivial assertion assuming ${\overline L} \, <
\,
1$ and  the ratio $\dc / \db$ is smaller than one. We now
prove the latter assumption. Remembering that $\dc$ is defined as $
{\sigma /( \pq -\ph)}$, and using Eqs. $\eosh$ and $\tzerottwo$ as well as
$\VTq$ and $\Latentheat$, this ratio satisfies
$$
{\dc \over \db } \, = \, {V(\Tq) \over \pq-\ph} \, < \, {V_c \over
\pq-\ph} \,\simeq \, {1 \over 6 \vq}
{ T^2_c - T^2_0 \over T^2_c } \, \leq \, 0.04\, . \eqn\dcdb
$$
To compute the parameter $\eta$, we only need to know the temperature
dependence of the interface velocity $\vb$, which
is\refmark{\dlhll,\lmt}
$$
\vb \, \sim \, \varepsilon \, = \,{T_c^2-\Tq^2 \over
T^2_c -T^2_0}\, . \eqn\vbepsilon
$$
The parameter $\eta$ [Eq.~$\etaterm$] then becomes
$$\eqalign{
\eta=\,-\Tq{d\vb\over dT}\vb\, &\simeq\, 2\, {\vb^2 \over
\varepsilon}\,{T^2_c
\over T^2_c-T^2_0} \cr
    &\simeq \, {\vb^2 \over \varepsilon}\, { 2 \pi \over \alpha_W}\, {m^2_H
    \over m^2_W}. \cr}
\eqn\etaEW
$$
As $\varepsilon$ behaves parametrically as $m_H^{-3/2}$,
we can obtain a lower bound for $\eta$ by considering the lowest value
for $m_H$, which, to be conservative, we take to be $40$~GeV, for which
$\varepsilon~\sim~0.25$\refmark{\dlhll,\lmt}. In such a
case,
$$
\eta \,  \gsim \, { 2 \pi \over \alpha_W} \, \vb^2 \eqn\etaalpha
$$
and unstable perturbations can only develop for
$$
\vb \, \lsim \, \sqrt{{\alpha_W \over 2 \pi}} \,\sim \, 0.07 \, , \eqn\vweak
$$
a range clearly not favored by the latest estimates: $ 0.1 < \vb <  0.9$.
{}From the numerical results above, we find that perturbations on all
scales are unable to destabilize the shape of a bubble:  if produced
spherical, it will stay spherical until it collides with other bubbles
and completes the transition. This conclusion, which contradicts a
previous analysis\refmark\freese, is illustrated in Figs. 6a and 6b.

For the QCD quark-hadron transition, the problem is entirely
nonperturbative and results are much less clear. ${\overline L}$ is
expected to be larger, in the range 0.5--1.
\unskip\break\noindent
In the cosmological context $v_b$
is rather small\refmark\keijo, of the order of 0.04. A fair calculation
of $\eta$ is at present an impossible task. From Eq.~$\etaEW$ we see
that unstable perturbations can only develop for
$$
\vb^2 \, < \, \epsilon \, \left( 1 - {T_0 \over T_c} \right)
= \,\, 1 - {\Tq \over
T_c} \, . \eqn\vbQCD
$$
The standard nucleation analysis for the cosmological
phase transition\refmark\keijo,
using values of $L$ and $\sigma$ measured with lattice Monte Carlo
techniques\refmark{\kkr-\potvin}, suggests
$1~-~\Tq/T_c~\approx~0.001$ or an upper limit on $\vb$ of about
$0.03$. This is
very close to the estimated value $0.04$ of $\vb$ and thus no definite
statement can be made. In terms of the present analysis, QCD, quite
interestingly, seems to be a borderline case.

\section{\bf Conclusions}

The prospect that baryogenesis took place at the electroweak scale has
recently been the motivation for the investigations of various aspects
of the electroweak phase transition (EWPT). In the present work, we
have analyzed the structure of a propagating front separating the
unbroken phase from the broken phase. In particular, we have
generalized the analysis of Landau\refmark\landau\ of the hydrodynamical
stability of a moving front under small perturbations. We have found
that, due to the particular nature of the EWPT, this generalization
leads to a conclusion opposite to previous works on the
subject\refmark\freese, namely, that {\it the growth of an electroweak
bubble is stable under small perturbations}.

This surprising result comes about because of the intrinsic weakness of
the EWPT, which is characterized by the smallness of the ratio of
scales ${ \dc / \db}\, $ ($ \sim \alpha_W$) and by the smallness of
the latent heat relative to the energy flowing across the interface
${\overline L}\,\sim \, \alpha_W^2 ( 90 / \pi^2g^{*})$, which,
in particular, result
in a large sensitivity $\eta \, \sim \, {1 / \alpha_W}$ of the
velocity of the interface on the temperature of the flowing
plasma. In general, there is a {\it critical velocity} below which
instabilities are allowed to develop. In the EW case, we found that
this critical velocity is parametrized as $ \sim\,
\sqrt{\alpha_W / 2 \pi}\, (m_W / m_H)$ and is bounded from
above by $0.07$. This value is significantly smaller than the
values obtained from microscopic
calculations\refmark{\dlhll,\lmt},
$0.1\, <\, \vb \, < \, 0.9$.

It is interesting to ask under what conditions a smaller value
of $\vb$ could be obtained. In the
minimal standard model, it would happen\refmark{\dlhll,\lmt}\ if the
mean free paths $\ell$ of the gauge bosons and of the top quarks are
larger than the thickness of the bubble wall, $\db$. The velocity
can in such a case reach a value of $0.04$ ($m_{top}~\sim~140$~GeV)
and is slightly smaller as the top quark mass
increases. In a non-minimal standard model, smaller values might be
allowed because of the slight dependence of $\vb$ on the microscopic
parameters.

In the QCD phase transition, too little is known to make definite
statements on the stability of the quark-hadron interface. One must
remember that even the order of the phase transition is not known at
present. It is quite interesting that reasonable estimates of the QCD
parameters invoked make the cosmological quark-hadron phase transition
a borderline case of our stability analysis.

The authors wish to thank the Institute for Theoretical Physics at
the University of California, Santa Barbara, where this work was begun.

\refout
\bigskip

{\bf Figure Captions}
\bigskip

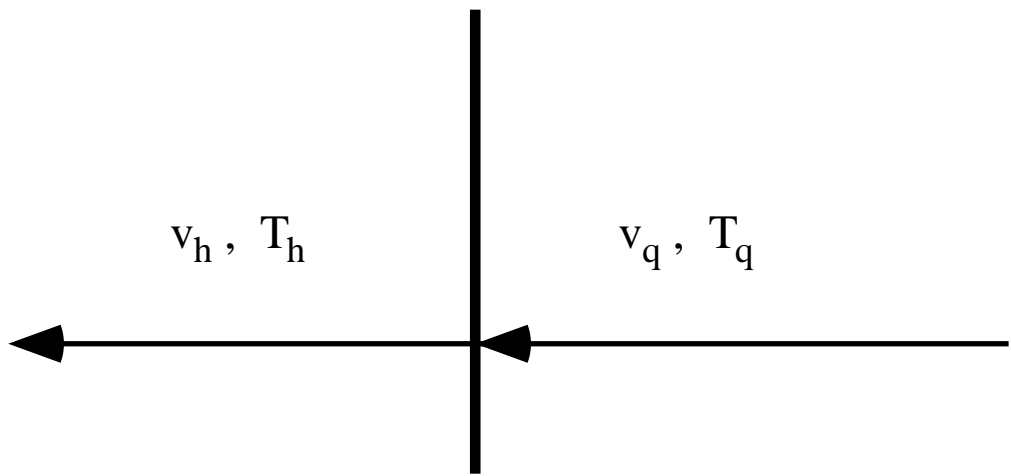
{\bf Figure 1}:  The burning front in the rest frame of the
front.  The wall is moving to the left, the inflowing
velocity of the high temperature phase is $\vq$ and the
outflowing velocity of the low temperature phase is $\vh$.
Deflagrations are when $\vq<\vh $ and detonations when
$\vq>\vh$.

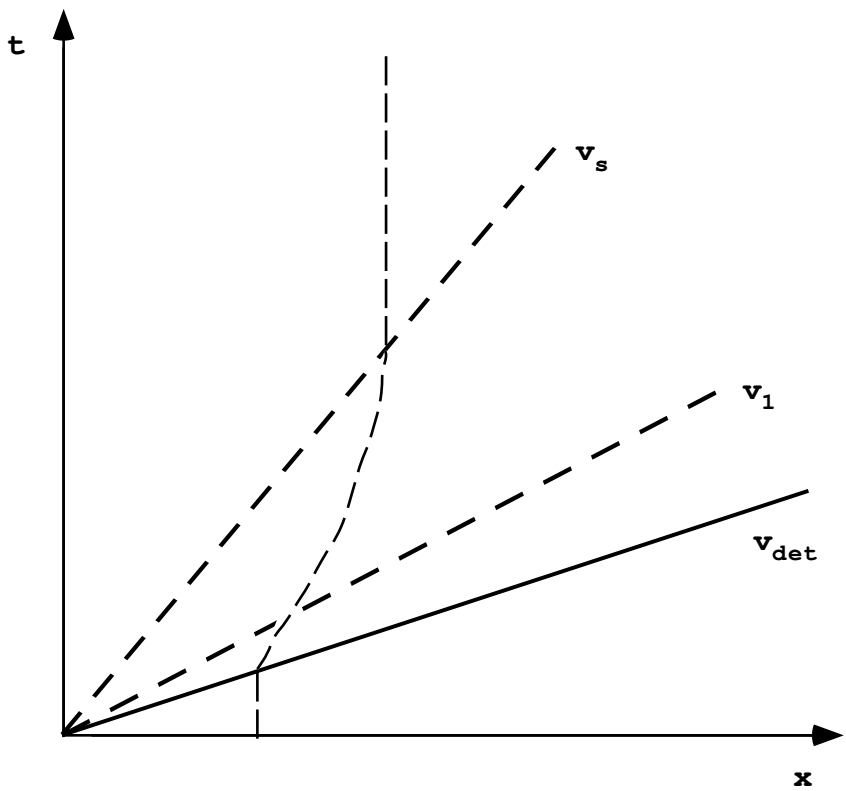
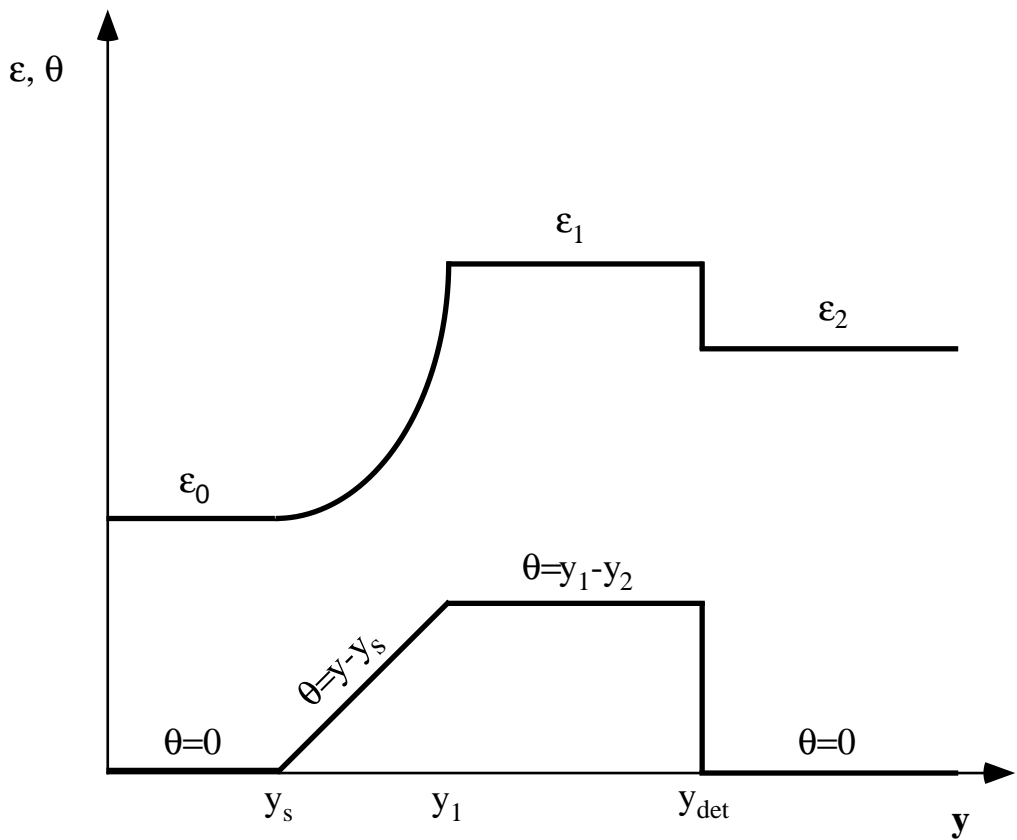
{\bf Figure 2}:  Structure of a detonation bubble in 1+1-dimensions:
(a) The time dependence of the $x>0$ part of the bubble with the three
fronts demanded by the boundary conditions. The dotted line shows
the path of a fluid element.
(b)  The energy density and flow rapidity of matter as functions
of space-time rapidity $y=\tanh^{-1}(x/t)$ for a detonation bubble.
For Jouguet detonations $y_1=y_s$ and no region of constant
velocity appears. The magnitudes of the quantities shown are
related by Eqs.~\epsy--\ezeroetwo.

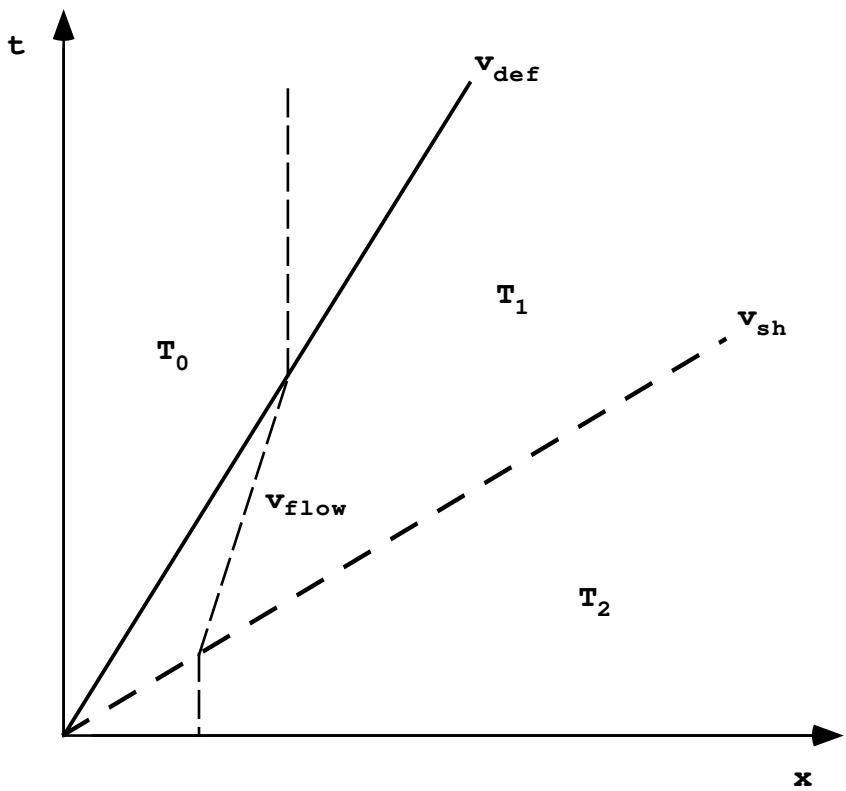
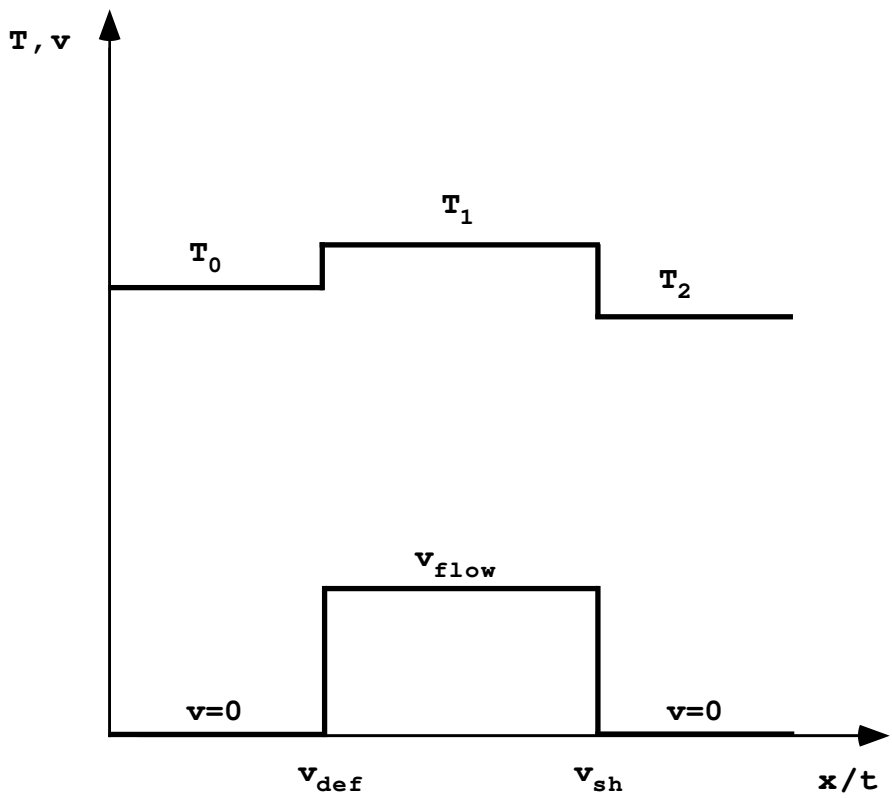
{\bf Figure 3}:  Structure of a deflagration bubble in 1+1-dimensions:
 (a) The time dependence of the $x>0$ part of the bubble with the two
fronts demanded by the boundary conditions. The dotted line shows
the path of a fluid element.
(b)  The energy density and flow rapidity of matter as functions
of $x/t$ for a deflagration bubble. The magnitudes of the quantities shown are
 related by Eqs.~\tzerottwo--\vflowdef.

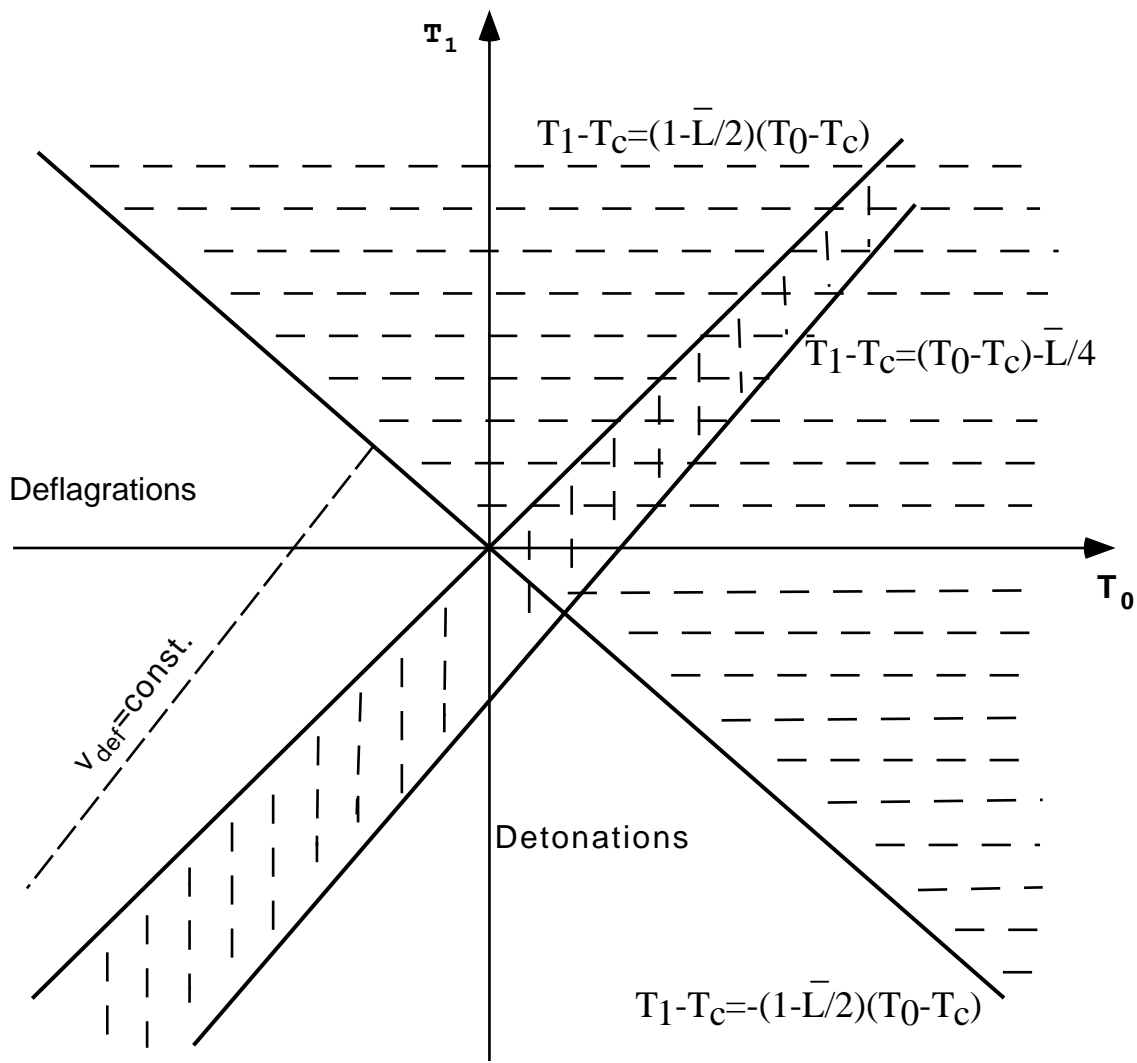
{\bf Figure 4}:  Regions allowed by the continuity of energy and
momentum fluxes and entropy increase for combustion from a high $T$
phase at $T=T_1$ to low $T$ phase at $T=T_0$. The three approximately
linear boundaries follow for $T_1-T_c,T_0-T_c\ll T_c$ from Eqs.~
\longequ\ by demanding the velocities to be real and from the second of
Eqs.~\sameas. The curves $v_{\rm def}$ = const. are explicitly given by
Eqs.~\tzerottwo\ and \tonettwo. Note that detonations are possible only
after supercooling by at least $\overline L/4$. Horizontal hatching
corresponds to the forbidden region characterized by $\Delta s < 0$,
and vertical hatching to imaginary velocity.

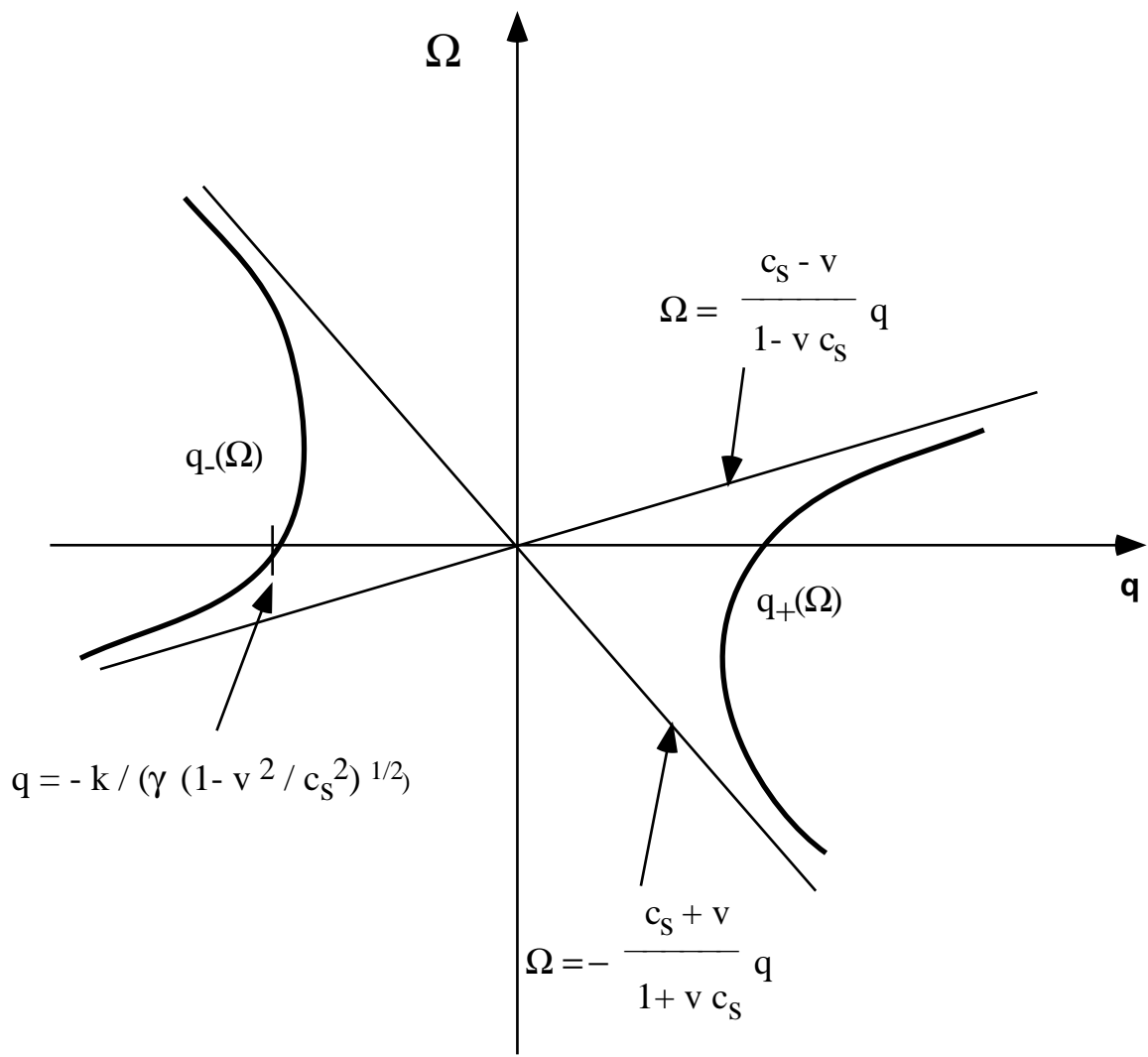
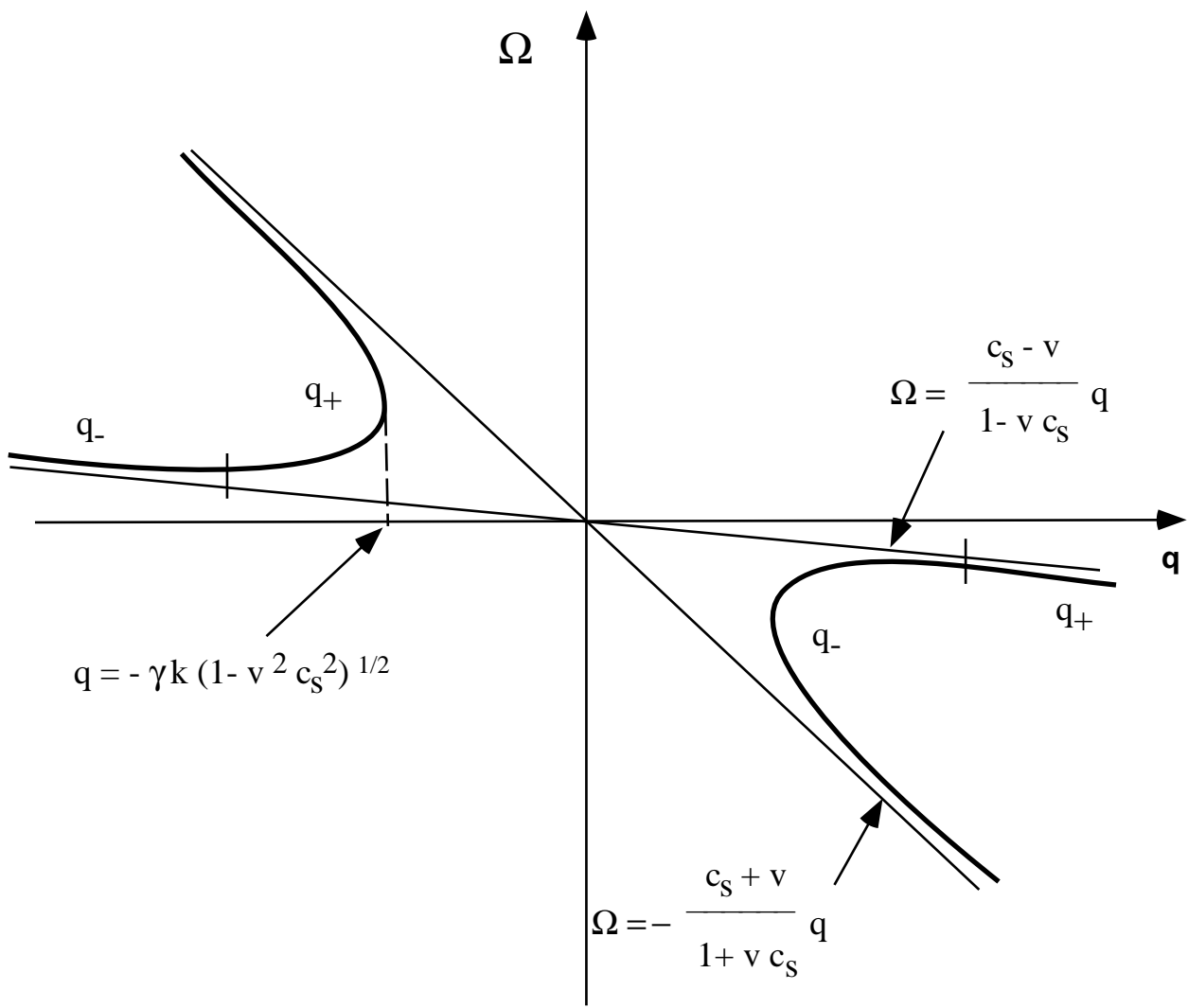
{\bf Figure 5}:  A plot of the dispersion relation $\qomegaplane$
for (a) $v<c_s$, (b) $v>c_s$.

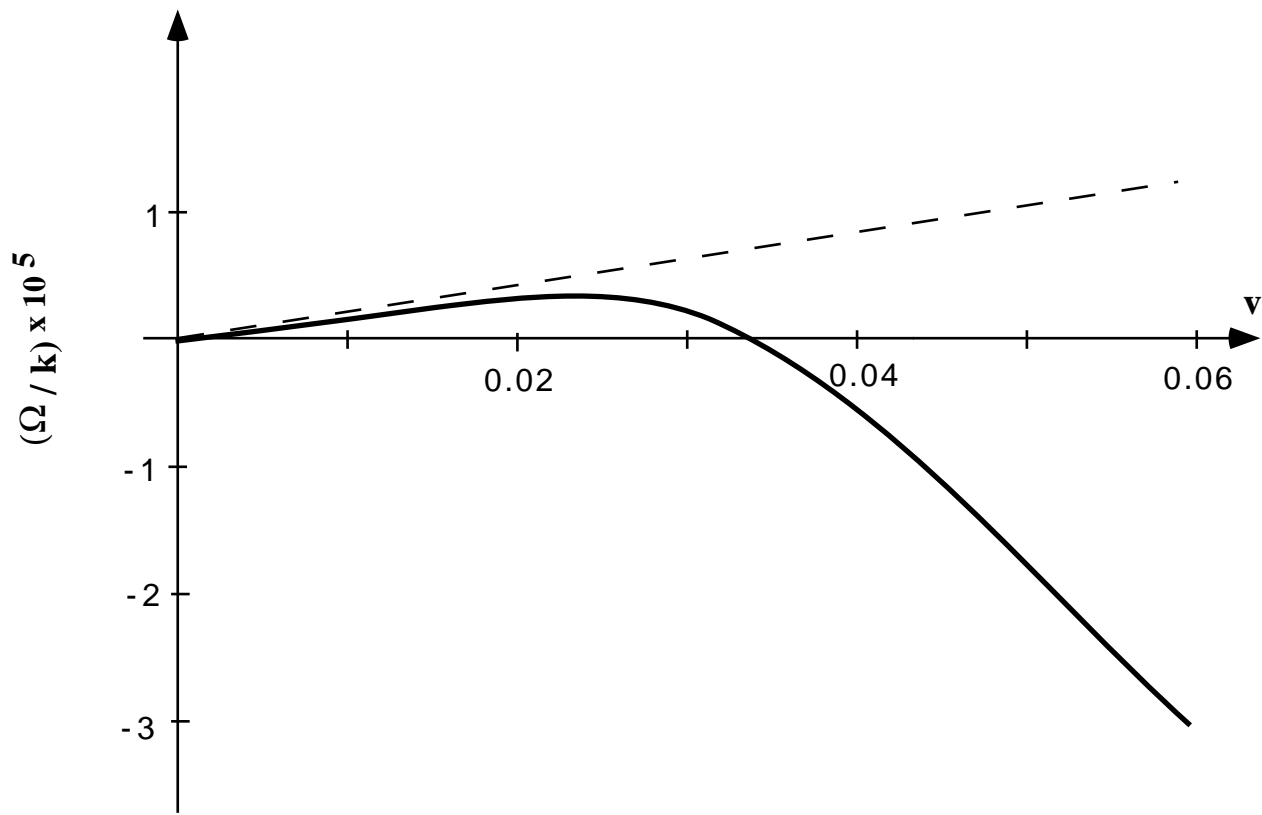
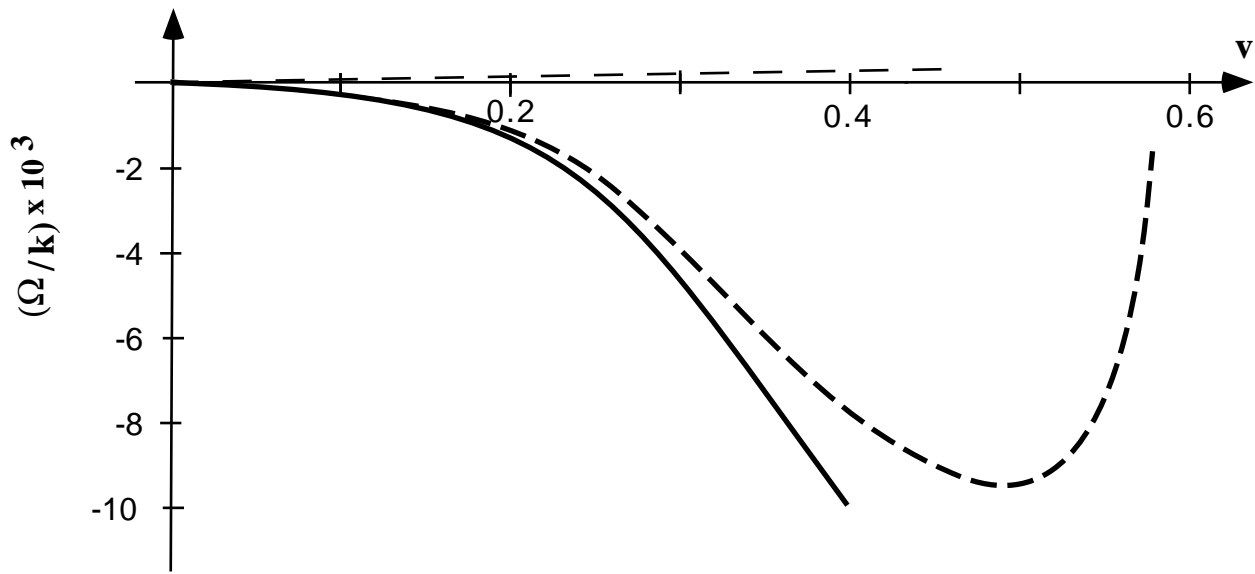
{\bf Figure 6}: Long wavelength limit of $\Omega/\k$ computed for
the electroweak theory ($m_h\,=\,60$ GeV). (a) Landau's results $\LL$
(dashed curve) and our results $\NRsol$ (solid curve) are shown for
small velocities $v \ll c_s$. (b) In the limit $v \rightarrow c_s$, the
inadequacy of $\LL$ and $\NRsol$ is illustrated by comparison with our
general formula $\Rsolution$ (dotted curve).

\bye